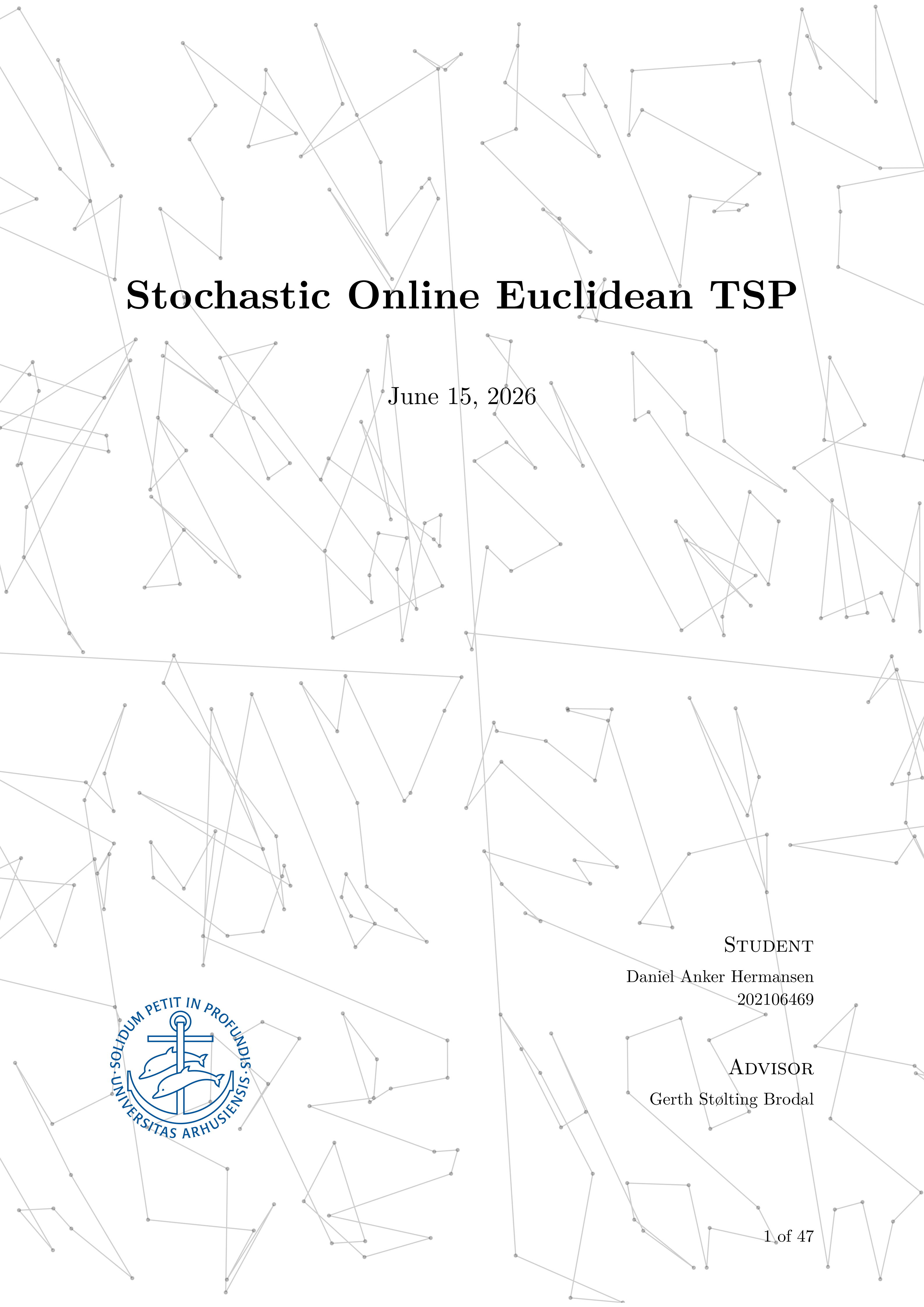

# Stochastic Online Euclidean TSP

June 15, 2026

STUDENT

Daniel Anker Hermansen
202106469

ADVISOR

Gerth Stølting Brodal

# Abstract

In the Euclidean travelling salesman problem (Euclidean TSP), a salesman must visit $n$ points in Euclidean space, while minimizing the travel distance, according to the Euclidean distance function. In online Euclidean TSP, introduced by Abrahamsen, Bercea, Beretta, Klausen and Kozma [ESA 2024], the points are revealed one at a time, and a time slot must be assigned before the next is revealed. Once a point is assigned to a time slot, it can never be reassigned to another time slot. There are $n$ time slots. Euclidean online TSP is a high-dimensional generalization of online sorting, introduced by Aamand, Abrahamsen, Beretta and Kleist [SODA 2023]. Bertram [ESA 2025] showed an algorithm that achieves a competitive ratio of $O(\sqrt{n})$ in the worst case. In stochastic online Euclidean TSP, the points are sampled uniformly and independently in the unit $d$-cube. Kalavas, Platanos and Tolias [STACS 2026] presented an algorithm achieving a competitive ratio of $O(\log^2 n)$ with high probability for stochastic online Euclidean TSP.

We present a simple algorithm that for $d \geq 2$ achieves an expected competitive ratio of $O(1)$, and for $d = 1$ achieves an expected competitive ratio of $O(\log n)$, matching the lower bound by Hu [SODA 2026] for $d = 1$. The algorithm is deterministic, and the expectation is due to the stochastic input. We also show that in the variant where there are more time slots than points, i.e., $\lceil (1+\varepsilon)n \rceil$ time slots and $d = 1$, our algorithm achieves an expected competitive ratio of $O(1 + \log \varepsilon^{-1})$. We also survey algorithms from the literature.

We experimentally evaluate our algorithm, which reveals that in all variants the constant factor hidden asymptotically is small. We also evaluate the algorithms from the literature.

# Resumé

I det euklidiske travelling salesman problem (euklidisk TSP), skal en sælger besøge $n$ punkter i det euklidiske rum, mens rejsedistancen minimeres, beregnet med den euklidiske distancefunktion. I online euklidisk TSP, introduceret af Abrahamsen, Bercea, Beretta, Klausen og Kozma [ESA 2024], bliver punkterne afsløret et punkt ad gangen, og før det næste punkt bliver afsløret, skal punktet tildeles et tidspunkt. Efter at et punkt har fået tildelt et tidspunkt, kan det aldrig tildeles et andet tidspunkt i stedet. Der er $n$ tidspunkter. Euklidisk online TSP er en højdimensionel generalisering af online sorting, introduceret af Aamand, Abrahamsen, Beretta og Kleist [SODA 2023]. Bertram [ESA 2025] præsenterede en algoritme som opnår en kompetitiv ratio på $O(\sqrt{n})$ i værste tilfælde. I stokastisk online euklididsk TSP bliver punkterne genereret tilfældigt, uniformt og uafhængigt i $d$-enhedskuben. Kalavas, Platanos og Tolias [STACS 2026] præsenterede en algoritme som opnår en kompetitiv ratio på $O(\log^2 n)$ med høj sandsynlighed for stokastisk online euklidisk TSP.

Vi præsenterer en simpel algoritme, som for $d \geq 2$ opnår en forventet kompetitiv ratio på $O(1)$, og for $d = 1$ opnår en forventet kompetitiv ratio på $O(\log n)$, hvilket passer med den nedre grænse af Hu [SODA 2026] for $d = 1$. Algoritmen er deterministisk og forventningen kommer fra det stokastiske input. Vi viser også at i varianten hvor der er flere tidspunkter end punkter, for eksempel $\lceil (1+\varepsilon)n \rceil$ tidspunkter og $d = 1$, opnår vores algoritme en forventet kompetitiv ratio på $O(1 + \log \varepsilon^{-1})$. Vi undersøger også algoritmer fra litteraturen.

Vi evaluerer vores algoritme eksperimentelt, hvilket viser, at I alle varianter er konstantfaktoren, som er skjult af asymptotikken, lille. Vi evaluerer også algoritmer fra litteraturen.

# Contents

# 1. Introduction

In the travelling salesman problem (TSP), you are given $n$ points and have to output a tour, i.e., an order in which the points shall be visited, such that the length of the tour is minimized. The length of the tour is the sum of the lengths of the segments between every adjacent point, including the segment from the last point to the first. The distance can be according to any metric, i.e., Euclidean.

In online Euclidean TSP, the points are revealed one by one in an online fashion and must be assigned a time slot, i.e, a position in the final order, before the next point is revealed. Let $a_i$ be the point placed in time slot $i$, then we define the cost as $\sum_{i=1}^{n-1} \|a_{i+1} - a_i\|_2$, i.e., the length of the path going through the points in the chosen order, according to the Euclidean distance function. The Euclidean distance of two points, in $d$ dimensions, $v$ and $u$ is defined $\|x - v\|_2 = \sqrt{\sum_{i=1}^{d} (x_i - v_i)^2}$. Note that this cost function is slightly different than the one normally used in TSP, as the segment from the last to the first point is not considered. However, the two costs are always within a factor of 2 of each other, due to the triangle inequality. Online TSP was introduced by Abrahamsen et al. [1]. All results in this thesis assume the Euclidean distance function, unless otherwise stated.

If $d = 1$, the points are simply real values, and the cost can be formulated as $\sum_{i=1}^{n-1} |a_{i+1} - a_i|$. This special case is called online sorting, due to the fact that if all points were known ahead of time, i.e., it was offline, the optimal solution would be to sort the points. In that case, the cost is the difference between the maximum and minimum points, which is also a lower bound. Online sorting was introduced by Aamand et al. [2] as a problem to show lower bounds and present algorithms for online packing of convex polygons in the plane. In online sorting, time slots are called cells. Online TSP is a later generalization of online sorting. This generalization naturally defines the cost function in online TSP to be the shortest path instead of the shortest tour.

The competitive ratio of an algorithm is the ratio between the cost of the output of the algorithm and the cost of the optimal solution in the offline version of the problem. The offline version of an online problem is where the entire input is known ahead of time. For Online TSP, the offline version is TSP, with a path instead of a tour. Therefore, the competitive ratio is the ratio between the output and the shortest (offline) path. The competitive ratio for Online TSP in all dimensions for any metric, satisfying the triangle inequality, not just the Euclidean distance function, is known to be $\Theta(\sqrt{n})$ [1, 2, 3] in the worst case.

A natural question to ask is if a better competitive ratio is possible in expectation if the points are chosen randomly instead of adversarially? In stochastic online TSP, the points are sampled uniformly from the $d$-dimensional unit cube. This variant was first considered in one dimension, where it is called stochastic online sorting. It was shown that an algorithm exists achieving an expected competitive ratio of $O\left((n \log n)^{\frac{1}{4}}\right)$ [1]. This was later improved to an expected competitive ratio of $\log n \cdot 2^{O(\log^* n)}$ and a lower bound of $\Omega(\log n)$ [4].

Another variant which has been considered is if there are more cells/time slots than points, say $m = \lceil (1 + \varepsilon)n \rceil$ cells/time slots. At the end, empty cells/time slots are ignored. In the adversarial case several results have been found [1, 2, 5, 6]. In this thesis, we only consider the stochastic variant, where in one dimension an algorithm achieving an expected competitive ratio of $O(1 + \varepsilon^{-1})$ was shown [1]. This bound is interesting as it is independent of $n$. To our knowledge, there are no known non-trivial results in the stochastic high-dimensional case.

### Our Results

We present a simple algorithm that improves the expected competitive ratio of stochastic online

Euclidean TSP both with and without extra time slots. The algorithm is exactly the same in all settings, where in the setting with extra time slots, we simply pretend that there are more points and then stop early. The algorithm need not even know how many points there will be, only the number of time slots. We first analyze the one-dimensional variant, stochastic online sorting, where we settle the expected complexity of the competitive ratio.

**Theorem 1.** *There exists an algorithm for stochastic online sorting achieving an expected competitive ratio of* $O(\log n)$.

We then show that if we have extra time slots, we get another bound based on $\varepsilon$.

**Theorem 2.** *There exists an algorithm for stochastic online sorting with* $\lceil (1+\varepsilon)n \rceil$ *cells achieving an expected competitive ratio of* $O(1+\log \varepsilon^{-1})$.

We then consider the high-dimensional case where, perhaps surprisingly, a better competitive ratio is possible. As there is a trivial lower bound of $\Omega(1)$, since no algorithm can produce a better solution than the optimal solution, this result is asymptotically optimal.

**Theorem 3.** *There exists an algorithm for stochastic online Euclidean TSP with* $d \geq 2$ *achieving an expected competitive ratio of* $O(1)$.

For completeness, we also consider the high-dimensional case with extra time slots, but since this is simply Theorem 3 , but the algorithm stops before all points are placed, the cost cannot be worse, and it is also optimal, as the trivial lower bound still applies.

**Corollary 4**: *There exists an algorithm for stochastic online Euclidean TSP with* $d \geq 2$ *and* $\lceil (1+\varepsilon)n \rceil$ *time slots, achieving an expected competitive ratio of* $O(1)$.

## 1.1. Terminology

We use the term point for a point in $d$-dimensional Euclidean space. A point can be thought of as a $d$-dimensional vector. For consistency, we use the term point in the one-dimensional case, i.e, in online sorting. We say that the algorithm outputs an array $A$ which is the order of the points. The array consists of cells. Therefore, when placing a point in a cell, this is the same as assigning a time slot to a point. A free cell is a cell in which no point is placed , and a filled cell is one where a point is placed. We use the term $d$-rectangle to describe the Cartesian product of $d$ intervals. Intuitively, this is simply an axis-aligned rectangle in $d$ dimensions. A point is inside a $d$-rectangle if it is contained in the Cartesian product. A $d$-cube is a $d$-rectangle with equal side length along all dimensions. A $d$-ball is a $d$-dimensional ball. A bucket is an ordered list of cells. The capacity of a bucket is the number of cells in the bucket. A bucket is full if all cells in it are filled. Buckets may have a $d$-rectangle associated with them. We call this a bucket's span. We say that a bucket spans a point if the point is inside the associated $d$-rectangle. We use $n$ to denote the number of points, $m$ the number of cells/time slots if it is not equal to $n$. We use $d$ to denote the number of dimensions.

## 1.2. Structure of the Thesis

In Section 2, we survey some algorithms and lower bounds from the literature. We focus on the techniques and the reasons why these techniques are not enough to reach optimality in the stochastic case. In Section 3, we present our simple algorithm. In Section 4, we prove Theorems 1, 2, and 3. We first prove properties about the algorithm which are used in the proofs of all three theorems, and then prove Theorem 1 in Section 4.1, Theorem 2 in Section 4.2, and Theorem 3 in Section 4.3. We evaluate some of the algorithms from the literature surveyed in Section 2, as well as our algorithm experimentally in Section 5. Finally, in Section 6, we conclude on the result and discuss future work.

# 2. Previous Work

We present some of the algorithms from previous papers. We focus on the techniques used and why they fail to obtain optimal algorithms in the stochastic case. In order to more easily address the algorithms presented, they are usually named based on the first author of the paper in which they were introduced. Note that this naming convention for the algorithms is exclusive to this thesis and not based on previous precedence. We start our survey with the worst-case algorithms. We call this the adversarial case, as the input sequence is decided by an adversary. The adversary knows where previous items have been placed.

## 2.1. Adversarial Online Sorting

A natural idea that one may come up with is to split the points into buckets akin to radix sort. To do this, split the universe into different areas, for example, intervals and partition points based on which interval contains the point. Additionally, split the output array $A$ into subarrays forming buckets and associate an interval with each. Placing a point into a bucket then actually means placing it in a free cell in that bucket. We call this technique bucketing.

Using the bucketing technique is not straightforward. We need to consider the possibility of overflow, i.e., when a bucket is not large enough to contain all points. We also need to consider the possibility of buckets never being filled, yet the cells must be filled by the algorithm eventually.

### 2.1.1. Aamand's Algorithm

Aamand et al. [2] shows how to use the bucketing technique to create an algorithm that achieves a worst-case cost of $O(\sqrt{n})$ if all points lie in the interval $[0,1]$. We call this algorithm Aamand's algorithm. Partition the universe, which is the interval $[0,1]$, into $\lfloor\sqrt{n}\rfloor$ equally-sized intervals. Partition the output array $A$ into $2\lfloor\sqrt{n}\rfloor$ equally-sized (up to rounding) buckets. Initially, the buckets do not have an interval associated with them, but they will later. When placing a point, first find any bucket that spans it and that is not full. If such a bucket is found, place the item in the first free cell of that bucket. Otherwise, choose any bucket with no interval associated with it yet, and associate the interval that contains the point with that bucket. If there are no such buckets, the algorithm recurses. Let $A'$ be the concatenation of all free cells of $A$. Then the algorithm recurses on $A'$.

Note that the algorithm only recurses when there are no empty buckets without an associated interval. That means that all buckets have an associated interval. Secondly, for any interval, there is at most one non-full bucket with the interval associated with it. Therefore, there are at most $\lfloor\sqrt{n}\rfloor$ non-full buckets, and each of these has at least one filled cell in it, meaning that at least half of the cells of $A$ have been filled. We therefore know that the recursion has size at most half of the current instance.

Let us count the cost of the algorithm. Inside a bucket, two points can be at most $\frac{1}{\lfloor\sqrt{n}\rfloor} = O\left(\frac{1}{\sqrt{n}}\right)$ far away from each other, so the cost between points inside buckets is at most $n \cdot O\left(\frac{1}{\sqrt{n}}\right) = O(\sqrt{n})$. Between two buckets, the cost can be at most 1, and there are $2\lfloor\sqrt{n}\rfloor - 1$ such boundaries contributing cost at most $O(\sqrt{n})$. Lastly, we have to consider the boundaries with the cells of $A'$. As all points are placed consecutively in each bucket, the part reused in $A'$ is also consecutive and therefore has at most 2 boundaries within each bucket. Since at most $\lfloor\sqrt{n}\rfloor$ buckets are not full, it contributes a cost of $O(\sqrt{n})$. For some constant $c$, we get the following recursion:

$$C(n) \leq C\left(\frac{n}{2}\right) + c\sqrt{n}, \tag{1}$$

which solves to $O(\sqrt{n})$. We make the assumption that $C(\cdot)$ is an increasing function. This is fine as the true worst case cost can be upper bounded by an increasing function.

The algorithm extends to the case where all points are in any interval, as long as that interval is known. If the interval is unknown, then the algorithm can be modified such that it learns the interval during the algorithm. This was shown by Abrahamsen et al. [1]. We still call it Aamand's algorithm even with this extension.

The key idea is to extend the interval whenever a point lands outside it. We use the first point as a center of the interval, meaning that the interval is always on the form $[x_1 - 2^k, x_1 + 2^k]$, where $k$ is an integer or $-\infty$, and is initially $-\infty$, meaning that the interval is $[x_1, x_1]$. Whenever a point is received which is outside the interval, $k$ is increased to the smallest $k$ such that it is inside the interval. Note that this changes the intervals associated with buckets, as these intervals are defined compared to the interval of the algorithm. For example, if there are 4 intervals, then the first is always $[x_1 - 2^k, x_1 - \frac{1}{2} \cdot 2^k]$, covering a forth of the full interval. In all other ways, the algorithm acts as before.

Let $\ell$ be the current length of the interval. All the analysis in regard to cost previously still holds up to a factor $2 \cdot 2^k \leq 2\ell$ except for between points in the same bucket, as points placed before and after an increase in $k$ respectively, can be further apart. Each bucket has at most one such boundary for each time $k$ is increased. The maximum difference here is $\ell$ at the time of the placement of the second point. Whenever $k$ is increased $\ell$ increases by at least a factor 2. Therefore, summing over all such boundaries in a bucket forms a geometric series and therefore the total extra cost is at most $2\ell$ per bucket and therefore $2 \cdot \ell \cdot 2 \cdot \lfloor\sqrt{n}\rfloor = \ell \cdot O(\sqrt{n})$. For some constant $c$, the recursion is therefore:

$$C(n) \leq C\left(\frac{n}{2}\right) + \ell \cdot c\sqrt{n}, \tag{2}$$

which solves to $\ell \cdot O(\sqrt{n})$. A slight detail is that $\ell$ may be different in each instance during the recursion. The $\ell$ in the solution is therefore the maximum $\ell$ over all instances.

Note that the optimal cost is at least the maximum of $\frac{\ell}{4}$ over all instances, as we know that $x_1$ is present and there is a value at least $\frac{\ell}{4}$ away from $x_1$ because otherwise the interval would not have been extended that far. Therefore, the algorithm achieves a competitive ratio of $O(\sqrt{n})$.

### 2.1.2. Lower Bound

Aamand's algorithm is optimal in the adversarial model. We now formally prove this. The proof is a slightly simplified version of the proof by Aamand et al. [2].

Let $k = \lfloor\sqrt{n}\rfloor$ and define the set $S = \left\{\frac{i}{k} \mid i \in \{0, 1, ..., k\}\right\}$. Our goal is to create a state where all points in $S$ are placed somewhere in $A$ such that each has an adjacent free cell. Then we only input 0, which causes the cost to be at least the sum of the elements in $S$, which is $\Omega(\sqrt{n})$. We do it as follows. If there is a point in $S$ where, everywhere it is placed in $A$, it has no adjacent free cell, then we input this point.

Notice that while we are inputting points from $S$, we know that the distance to the closest neighbor is at least $\frac{1}{k}$. Therefore, if we end up filling $A$ entirely with this method, then all neighbors have distance at least $\frac{1}{k}$, and therefore the cost is $\Omega(\sqrt{n})$. On the other hand, if at some point there is no such point from $S$, we input 0 for the rest which, as described earlier, results in $\Omega(\sqrt{n})$ cost as well.

As all points are in $[0, 1]$, the optimal cost of the input is at most 1, and therefore the competitive ratio is at least $\Omega(\sqrt{n})$.

## 2.2. Adversarial Online TSP

Attempting to extend Abrahamsen's algorithm into higher dimensions reveals some of the challenges of higher dimensions. Instead of using intervals to split the points into buckets, it may be natural to use either $d$-rectangles, $d$-cubes or $d$-balls. All options have the problem that the number required to cover the universe is a factor $2^d$ larger than in one dimension if the side lengths are the same, and otherwise the side length must be much larger. For $d$-balls, there is the additional challenge that $d$-balls do not partition Euclidean space.

### 2.2.1. Abrahamsen's TSP algorithm

Abrahamsen et al. [1] used the idea with $d$-cubes to construct an algorithm, which we call Abrahamsen's TSP algorithm. They presented it, ignoring the dependence on $d$, in which it has a competitive ratio of $O(\sqrt{n \log n})$. However, including the dependence on $d$ results in a competitive ratio of $O\left(2^d \sqrt{dn \log n}\right)$.

The key idea of the algorithm is that some of the $d$-cubes will be empty, and if many are filled, the optimal cost in the offline version also gets larger. In the presentation of the algorithm, it is assumed that all points are in $[0, 1]^d$. Using a doubling construction, where the problem is restarted on the remaining cells with a larger universe, the analysis generalizes.

Let $K$ be the number of $d$-cubes which contains at least one point of the input sequence. As a subprocedure, they use an algorithm for the uniform metric online TSP that achieves a competitive ratio of $O(\log n)$. The uniform metric here means that for two equal points, the distance is 0, and otherwise the distance is 1. We consider all points in the same $d$-cube to be equal. Suppose the $d$-cubes have side length $\ell$, then the distance between points in the same $d$-cube is at most $\sqrt{d} \cdot \ell$ and the distance between points from different $d$-cubes is at most $\sqrt{d}$. However, we know that there are only $O(K \log n)$ such neighbours in $A$. This, together with neighbors inside the $d$-cubes, yields a cost of $\sqrt{d} K \log n + \ell \cdot \sqrt{d} n$. Consider the optimal path visiting all $K$ $d$-cubes in some order. Among every $2^d + 1$ consecutive $d$-cubes, at least two are not neighbors, by the pigeon-hole principle, and therefore have distance at least $\ell$. Therefore, the optimal offline cost is at least $\frac{K \cdot \ell}{2^d + 1}$. Setting $\ell = \sqrt{\frac{\log n}{n}}$ yields a competitive ratio of $O\left(2^d \sqrt{dn \log n}\right)$.

All that remains is to build the algorithm for the uniform metric, and it is very simple. When placing a point, it is placed in the cell immediately following where the last point equal to this point was placed. If that cell is filled or there have not yet been any points equal to this point, the middle cell of the largest consecutive free subinterval is chosen. A cost of one is incurred in the latter case, while no cost occurs in the former, so we only need to bound how many times the latter can happen. Note that there can be at most $K + 1$ maximally consecutive free intervals, as a subinterval is split only when introducing a new point not yet seen or that point previously filled a cell where both neighbours were filled, thereby reducing the number of maximally contiguous intervals. Suppose that at some point the largest maximally contigous interval has length $t$, then after $K$ splits, all have length at most $\frac{t}{2}$. These halving happen at most $\log n$ times, therefore the competitive ratio of $O(\log n)$, since the optimal cost is $K - 1$.

The result is quite bad for large dimensions. Part of the reason may be that the optimal cost is not very well estimated. The lower bound estimate is very pessimistic. However, it is not clear how to make a less pessimistic estimate.

### 2.2.2. Bertram's Algorithm

Bertram [3] instead considered the idea of using $d$-balls, and instead of packing these $d$-balls in space, they are instead placed greedily as needed. The algorithm makes use of the minimum spanning tree to set parameters during the running of the algorithm, which ties these parameters to the optimal offline path since the minimum spanning tree is constant approximation of TSP [7]. We call this algorithm Bertram's algorithm.

For simplicity, suppose we knew the optimal path length $L$, then by setting $r = \frac{L}{\sqrt{n}}$, we know that $\sqrt{n}$ $d$-balls can be placed such that all points are contained by at least one ball. This can be done by covering the path in balls by setting the first ball's center at the first point of the path, and then the next at the first point in the path which is not covered by the first ball, and so on. Additionally, this can be done greedily by taking the points in any order, and if no ball covers it, place a ball centered there. Using Aamand's algorithm but with these balls instead of intervals, where the balls are defined when we receive a point which is not covered by a ball, we get that the distance between two points in the same ball is at most $2r = 2\frac{L}{\sqrt{n}}$. The distance between two balls is at most $L$ by the triangle inequality, and therefore the overall cost is $O(\sqrt{n}L)$, using the same analysis as for Aamand's algorithm. This results in a competitive ratio of $O(\sqrt{n})$.

Of course, we do not know $L$, so we estimate it. By finding the minimum spanning tree $M$ of the points seen so far, we get a value which is at least half of $L$. Let $r = \frac{4M}{\sqrt{n}}$. When a point is not contained in a ball, and we already have $\sqrt{n}$ balls, we restart with a new $r$. Since $r \geq \frac{2L}{\sqrt{n}}$ at the time of definition, when restarting, we know that $L$ must have doubled for each restart. As in the analysis for Aamand's algorithm, the competitive ratio is therefore $O(\sqrt{n})$.

The analysis only requires that the MST is a 2-approximation and that $\sqrt{n}$ $d$-balls can cover $n$ points greedily. Both of these are true for any metric that satisfies the triangle inequality. This result therefore holds, not only for the Euclidean metric.

## 2.3. Stochastic Online Sorting

A challenge we had with the bucketing technique is that we do not know how many points each bucket should contain. This formed the problem of bucket overflow and cells left over. When the input is generated stochastically uniformly independent at random, we still do not know, but we have high probability bounds. We can use this to make buckets have just the right capacity such that with high probability it is filled, or we can create them large enough such that with high probability the bucket does not overflow. This means that we can eliminate one of the issues. Recall that in this version, we know that all points are in the interval $[0, 1]$.

### 2.3.1. Abrahamsen's Algorithm

We can use this idea to create an algorithm that beats Aamand's algorithm when the input is uniformly random. This was shown by Abrahamsen et al. [1]. We split the interval $[0, 1]$ into $t$ intervals. We then create $t$ buckets from $A$ such that each bucket's capacity is maximized while it is eventually filled with probability at least $1 - \frac{1}{n^2}$. The remaining cells are called the backyard. When receiving a point, we place it into the bucket that spans the point. If the bucket is full, it is instead placed into the backyard.

But how are the points placed inside the buckets? Note that the points which are placed in the buckets are uniformly distributed in a subinterval. We can therefore recurse on a scaled version in the buckets. For the backyard, we can use Aamand's algorithm. In case we are unlucky and at least one of the buckets is not filled, when the backyard overflows, we just place the remaining points anywhere. By a union bound, the probability of this is at most $\frac{1}{n}$. Therefore, in case it fails, the competitive ratio is trivially bounded by $n - 1$, and therefore it only contributes 1 to the expected competitive ratio. If the algorithm succeeds, it becomes a recursion. Setting the capacities appropriately, we get an expected competitive ratio of $O\left((n \log n)^{\frac{1}{4}}\right)$. Note that the algorithm succeeding does not necessarily mean that the recursion succeeds. This is, however, not a problem due to the linearity of expectation.

While this is better than the competitive ratio in the adversarial case, it is not much better and still polynomial. A big issue with this approach is the capacity of the backyard. Using an adversarial algorithm for the backyard, therefore, always results in a high cost.

### 2.3.2. Hu's Algorithms

Fortunately, we can remedy the issues of the backyard by noting that the points in the backyard still have some structure, which we can use. The general idea is that if we look at a fixed point in time, some of the buckets are filled and some are not. So the next point is either placed in one of the unfilled buckets, or it is placed in the backyard. But if it is placed in the backyard, we know that it was uniformly distributed among the intervals of the buckets that are filled. At the time when the first bucket is full, most other buckets are likely to be somewhat filled. At this point, we have $n'$ free cells left. We can therefore imagine that we have a completely new instance with $n'$ cells. Then calculate the maximimum capacity of the buckets where they are filled with high probability. The key idea is that some of these $n'$ cells are part of the buckets and some of the backyard. We can therefore imagine extending the bucket into the backyard such that it has the right capacity. Really, they are going to be two buckets, and we first fill the original bucket and then the bucket made from cells from the backyard. We call the technique of creating buckets in multiple phases the phases technique. We say that a new phase has started when a new set of buckets is allocated. We call the technique of allocating buckets with different capacities based on previous input adaptive allocation.

As an example, imagine if we have $n = 100$ and 4 buckets, we may choose to make buckets of capacity 20 and a backyard of capacity 20. When the first bucket is full, we have $4, 5, 0$ and 4 free cells in the four buckets, respectively. Therefore, $n' = 33$. We may choose to make buckets of capacity 6 and a backyard of capacity 9. We then allocte buckets of capacities $2, 1, 6$ and 2 in the backyard. Therefore, in total, there are 6 cells assigned to each interval. Note that the numbers are completely made up and do not necessarily satisfy the high probability requirement.

To use this technique, we must ensure that all buckets are filled with high probability and that buckets can be allocated with high probability. We may end up in a situation where the number of cells needed from the backyard is negative, in which case the technique fails. But if this is unlikely, we can still get a good expected competitive ratio. We can simply place the remaining points wherever, since the competitive ratio can never exceed $n - 1$.

Hu [4] used this idea to create an algorithm that achieves an expected subpolynomial competitive ratio. We call this algorithm Hu's adaptive algorithm. Choose the number of buckets. It works for all choices at least 2, so we choose 2. Then allocate two buckets and a backyard, where the backyard has capacity $n^{\frac{2}{3}}$, and the two buckets have equal capacity. Recurse in the buckets. When one bucket is full, adaptively allocate new buckets as described before. Once the remaining items are few, just place the rest in the backyard wherever. The cost can be seen as a recursion. The two buckets have a capacity of at most $\frac{n}{2}$, so we can use that as an upper bound. In each recursion, the universe is half the size, so the cost is also halved. With high probability, the number of remaining free cells, when one bucket is full, can be shown to be $\Theta\left(n^{\frac{2}{3}}\right) \le c' n^{\frac{2}{3}}$, for some $c'$ and large enough $n$. Additionally, there are two gaps between the buckets and the backyard, which cost at most 2. We also get an extra cost of 1 in expectation for the case where the algorithm fails. Assuming $C(\cdot)$ is increasing, we get the following recursion as an upper bound:

$$C(n) \le C\left(\frac{n}{2}\right) + C\left(c' n^{\frac{2}{3}}\right) + 3. \tag{3}$$

Therefore, by induction $C(n) \le cn^{\varepsilon}$ for any $\varepsilon > 0$ and some constant $c > 0$ depending on $\varepsilon$.

$$\begin{aligned} C(n) &\le c\frac{n^{\varepsilon}}{2^{\varepsilon}} + c \cdot c'^{\varepsilon} \cdot n^{\varepsilon\frac{2}{3}} + 3 \\ &\le c\left(\frac{n^{\varepsilon}}{2^{\varepsilon}} + 2 \cdot c'^{\varepsilon} \cdot n^{\varepsilon\frac{2}{3}}\right) \\ &\le cn^{\varepsilon}, \end{aligned} \tag{4}$$

for large enough $n$. The math follows from the fact that for any constant $k$: $n^{\frac{2}{3}} \le kn$ for large enough $n$, where $n$ is chosen depending on $k$. This proof is rather unsatisfactory, as it does not give us any insight into how good the algorithm actually is, but it does prove that subpolynomial is possible in expectation.

Suppose the true cost could be written on the form $(\log n)^{f(n)}$. What could $f(n)$ be? We know that if $f(n) = O(\log \log n)$, then it is subpolynomial. We therefore may try to see if it solves the recursion and therefore is an upper bound. We have not been successful, but cannot rule it out either. We look into this more in the evaluation.

Hu also introduced the technique of merging. The idea is that the number of buckets need not be the same in every phase. If we reduce the number of buckets by some factor between two phases, then when adaptively allocating, we take into account all buckets whose span is a subspan of the new bucket. Hu's merging algorithm is based on this technique. It chooses some number of buckets to begin with and a merging factor $K = \Theta(\text{poly}\log(n))$. The idea is to make every bucket have capacity $\Theta(\text{poly}\log(n))$, while ensuring that all buckets are allocatable with high probability. Between some phases, we merge buckets $K$-wise if the buckets would otherwise become too small. Within each bucket, points are placed wherever. Hu showed that with a parameter set, this does not fail with high probability. This results in a competitive ratio of $O\left(\frac{\log^{21} n}{\log \log n}\right)$ with high probability.

Hu's final algorithm combines the techniques of adaptive allocation, merging, and recursion in one. The recursion is tricky, as not all buckets get uniformly distributed points. However, the same ideas as for the adaptive allocation can be used, by giving the information of how many cells are free in each subinterval of the interval that the recursion spans. Hu showed that parameters exist such that the algorithm succeeds with high probability. An important thing is that the merging factor in all recursions have to be related as the number of buckets in a phase must be a power of the merging factor, as otherwise we do not know how to adaptively allocate. Secondly, we must ensure that it does not fail with high probability. Hu showed that such parameters do exist where buckets have $\Theta(\text{poly}\log n)$ capacity. Let $K$ be the merging factor. The buckets are set to have a maximum capacity of $K^{2.5}$.

First, the number of phases is lower bounded by the number of times we can merge, since we must end with one bucket in the last phase. As $K = \log^{c'} n$ for some constant $c'$, the number of phases is at least $\frac{\log n}{c' \cdot \log \log n}$ for large enough $n$. It can also be shown that the number of phases is actually $O\left(\frac{\log n}{\log\log n}\right)$, since merging happens often. Therefore, for some constant $c'' \geq 1$, the number of phases is bounded by $c'' \cdot \frac{\log n}{c' \log\log n}$. In each phase, there are a number of buckets, but adding up the maximum difference on their boundaries, you get 2. Additionally, there is a boundary with the next phase, adding 1 cost. There is a number of recursions of capacity at most $\log^{2.5 \cdot c'} n$. But each of these is a scaled version according to how many buckets there are. This creates the following recursion:

$$C(n) \leq \left(C\left(\log^{2.5 \cdot c'} n\right) + 3\right) \cdot c'' \cdot \frac{\log n}{c' \cdot \log\log n}. \tag{5}$$

If you try to solve this recursion with $C(n) \leq c \log n$, you will fail, but note that you fail by some constant factor, no matter the true unknown value of $c''$.

$$\begin{aligned} C(n) &\leq c \cdot (2.5 \cdot c' \log\log n + 3) \cdot c'' \cdot \frac{\log n}{c' \cdot \log\log n} \\ &\leq 4 \cdot c \cdot c' \log\log n \cdot c'' \cdot \frac{\log n}{c' \cdot \log\log n} \\ &= 4 \cdot c'' \cdot c \cdot \log n. \end{aligned} \tag{6}$$

Therefore, it can be solved by multiplying a constant factor in each recursion. Since the recursion depth is $O(\log^* n)$, since each recursion has $\operatorname{poly}\log(n)$-size, the recursion is solved by $C(n) \leq \log n \cdot 2^{O(\log^* n)}$. It is due to the non-constant recursion depth that this algorithm fails to have a competitive ratio of $O(\log n)$.

All of Hu's algorithm, while theoretically interesting are very complicated when working out all the details and not at all practical. Hu's merging algorithm and Hu's final algorithm use parameters in the order of $\log^8 n$ and $\log^{16} n$ respectively, which makes them impossible to evaluate on practical hardware, as they would just collapse into a much simpler algorithm immediately, as these parameters are huge. For this reason, we do not experimentally evaluate these two algorithms.

All the bounds in Hu's algorithm are based on Chernoff bounds. A question one might ask is if there could exist parameters where the competitive ratio is $O(\log n)$? We have tried, but to no avail. It seems like an impossibility to balance the merging factor and the bucket capacity to make it work. The bucket capacities seem to have to be $\Omega(\operatorname{poly}\log n)$, which causes non-constant recursion depth and the merging factor seems to have to be smaller than the bucket capacity, which makes the recursion impossible to solve with $O(\log n)$. In fact, in the Open Questions section of [4], Hu writes that he believes that $O(\log n)$ is impossible due to inherent drawbacks in Chernoff bounds, and that this can be used to prove an $\omega(\log n)$ lower bound for stochastic online sorting in general. The latter is wrong, as we prove in Section 4.1, but it does seem that the techniques we have discussed so far are not enough to achieve $O(\log n)$.

#### 2.3.3. Kalavas's Algorithm

Concurrently to Hu, Kalavas et al. [8] studied the problem and found the same techniques as Hu. Kalavas et al. decided to choose parameters quite differently, making their algorithm most comparable to Hu's merging algorithm. The key differences are that the merging factor is always 2, which ensures that the number of phases is $\Theta(\log n)$. Secondly, the buckets have capacity $\Theta(\log^2 n)$. Similar to Hu's merging algorithm, recursion is not used; instead, Aamand's algorithm is used to place points within each bucket. Since the capacity is $\Theta(\log^2 n)$ and Aamand's algorithm has a competitive ratio of $O(\sqrt{n})$, the cost per phase is $O(\log n)$. This gives an overall cost of $O(\log^2 n)$. Practically, the output array $A$ is split into buckets with some capacity $B = \Theta(\log^2 n)$. which is used to split $A$ among the phases. The actual buckets in each phase may have larger or smaller capacity due to adaptive allocation, but still $\Theta(\log^2 n)$.

They proved that the buckets are filled with high probability before the next phase is finished, meaning that only two sets of buckets can be active at any point. This makes the algorithm much more practical than Hu's algorithms. Additionally, because of no recursion, and the algorithm to place points in each bucket is infallible, the algorithm achieves the competitive ratio with high probability. They state that the algorithm can be improved to a competitive ratio of $O\left(\log^{\frac{3}{2}+\varepsilon} n\right)$ for any $\varepsilon > 0$ by tweaking the parameters.

We now prove that the algorithm succeeds with high probability. We present a simpler proof than the one presented in [8]. For simplicity, we ignore all rounding.

We set up two conditions for each phase. If these two conditions hold for all phases, then the algorithm definitely succeeds. Terminology-wise, we say that each phase has a number of spans, and at most three buckets in that span, which is one from this phase and two from the previous phase. Let $B = c \ln^2 n$, for some constant $c$.

① When the first span is full, then all other spans are filled at least by a factor $\frac{5}{6}$.
② Each span from the previous phase gets filled with at least $\frac{1}{4}B$ points during this phase.

We derive some facts from these conditions. The capacity of each span of a phase is at least $B$ and at most $\frac{3}{2}B$. Let there be $k$ spans in this phase. At least $kB$ cells are initially allocated for this phase, so the capacity of each span is at least $B$. For the upper bound, we use induction. For phase 1, there are no cells from the previous phase, so the capacity is always $B$. For all other phases, since each span is filled by a factor $\frac{5}{6}$ due to ①, the number of cells remaining is at most $\frac{1}{6} \cdot \frac{3}{2}B = \frac{1}{4}B$. Since there are $2k$ such spans in the previous phase, the number of cells in the current phase is at most $\frac{1}{4}B \cdot 2k + kB = \frac{3}{2}kB$. Since there are at most $\frac{1}{4}B$ remaining cells in each of the previous phase's spans and ②, all leftover cells from the previous phase are filled during this phase. Therefore, these two conditions are sufficient for the algorithm to succeed.

We bound the probability of failure. The probability for a phase to follow the conditions depends on previous phases also following the conditions. We use a union bound over all phases. We can do this as the dependencies on the events are non-cyclic.

let $kB \leq L \leq \frac{3}{2}kB$ be the capcity of this phase. We show that after $\frac{11}{12}L$ points have been placed, the following three conditions hold with high probability.

① All spans in this phase have received at least $\frac{5}{6}\frac{L}{k}$ points.
② All spans in this phase have received at most $\frac{L}{k}$ points.
③ At least $\frac{1}{4}B$ points have been received in each of the previous phase's spans.

Note that these three conditions imply the two previous conditions. These are simply more specific conditions, as we say when exactly the conditions must be true. We show all three with Chernoff bounds.

We use the following Chernoff upper bound [9, Corollary 4.6] for a binomial distribution $X$ with mean $\mu$:

$$\Pr[X \geq (1+\delta)\mu] \leq e^{-\frac{\delta^2 \mu}{3}}. \quad (7)$$

We use the following Chernoff lower bound [9, Theorem 4.5] for a binomial distribution $X$ with mean $\mu$:

$$\Pr[X \leq (1-\delta)\mu] \leq e^{-\frac{\delta^2\mu}{2}}. \tag{8}$$

For ① there are $k$ spans, therfore $\mu = \frac{11}{12}\frac{L}{k} = \frac{11}{12}B$. We set $\delta = \frac{1}{11}$. Let $X$ be a random variable denoting the number of points received in the span.

$$\begin{aligned} \Pr\left[X \leq \frac{5}{6}\frac{L}{k}\right] &\leq \Pr[X \leq (1-\delta)\mu] \\ &= e^{-\frac{\delta^2\mu}{2}} \\ &\leq e^{-\frac{1 \cdot 11B}{2 \cdot 11^2 \cdot 12}} \\ &= e^{-\frac{11c\ln^2 n}{2 \cdot 11^2 \cdot 12}} \\ &= n^{-\frac{11c\ln n}{2 \cdot 11^2 \cdot 12}}. \end{aligned} \tag{9}$$

For ② there are $k$ spans, therfore $\mu = \frac{11}{12}\frac{L}{k} = \frac{11}{12}B$. We set $\delta = \frac{1}{11}$. Let $X$ be a random variable denoting the number of points received in the span.

$$\begin{aligned} \Pr\left[X \geq \frac{L}{k}\right] &\leq \Pr[X \geq (1+\delta)\mu] \\ &= e^{-\frac{\delta^2\mu}{3}} \\ &\leq e^{-\frac{1 \cdot 11B}{3 \cdot 11^2 \cdot 12}} \\ &= e^{-\frac{11c\ln^2 n}{3 \cdot 11^2 \cdot 12}} \\ &= n^{-\frac{11c\ln n}{3 \cdot 11^2 \cdot 12}}. \end{aligned} \tag{10}$$

For ③ there are $2k$ spans, therefore $\mu = \frac{11}{12}\frac{L}{2k}$ and $\frac{11}{24}B \leq \mu \leq \frac{11}{16}B$. We set $\delta = \frac{7}{11}$. Let X be a random variable denoting the number of points in the span.

$$\begin{aligned} \Pr\left[X \leq \frac{B}{4}\right] &\leq \Pr[X \leq (1-\delta)\mu] \\ &= e^{-\frac{\delta^2\mu}{2}} \\ &\leq e^{-\frac{7^2 \cdot 11B}{2 \cdot 11^2 \cdot 24}} \\ &= e^{-\frac{7^2 \cdot 11c\ln^2 n}{2 \cdot 11^2 \cdot 24}} \\ &= n^{-\frac{7^2 \cdot 11c\ln n}{2 \cdot 11^2 \cdot 24}}. \end{aligned} \tag{11}$$

If we set $c$ sufficiently high, i.e., $c = 800$ the probability in (9) , (10) and (11) is at most $\frac{1}{n^2}$. Using a union bound over every span of which there are $k$ in (9) , $k$ in (10) , and $2k$ in (11) , we get that it fails in the phase with probability at most $\frac{4k}{n^2}$. As the sum over $k$ over all phases is at most $n$, as there cannot be more than $n$ buckets, the probability of failure is at most $\frac{4}{n}$. An arbitrarily small probability can be chosen by choosing a larger $c$.

You may notice that all bounds still hold if the capacity of the buckets is only $\Theta(\log n)$. In that case, the cost is be $O\left(\log^{\frac{3}{2}} n\right)$, which is better than what they state they can get by tweaking parameters.

#### 2.3.4. Lower Bound

After having placed $n - i$ points, there are only $i$ free cells remaining. If we assume that all placed items are individually uniformly random, then placing an item would yield an expected $\Omega\left(\frac{1}{i}\right)$ extra cost. This would mean a cost of at least $\Omega(\log n)$. Hu [4] showed that there, in fact, is such a lower bound, based on that intuition. We show a proof here, inspired by the original proof, but modified to make less use of asymptotic notation.

Hu's proof is for any randomized algorithm, but as the input is chosen independent of the random choices of the algorithm, allowing randomization is rather meaningless. Any randomized algorithm is a distribution over deterministic ones. And therefore, the lower bound must also hold for the deterministic algorithm with the lowest expected cost. It therefore suffices to show it for deterministic algorithms.

We focus on pairs of adjacent cells. Let $\{x_n\} = \{x_1, x_2, ..., x_n\}$ be the input to the algorithm and $a_1, a_2, ..., a_n$ be the output of the algorithm.

Let $T_j$ be a random variable denoting when the $j$'th cell is filled. Note that for an integer $1 \leq i \leq n$ and $1 \leq j < n$:

$$\Pr_{\{x_n\},j}\left[T_j > n-i \lor T_{j+1} > n-i\right] \geq \max\left\{\Pr_j\left[T_j > n-i\right], \Pr_j\left[T_{j+1} > n-i\right]\right\} \geq \frac{i}{n} \quad (12)$$

as $i$ of all $T_j$ are greater than $n-i$. Conditioned on $T_j > n-i \lor T_{j+1} > n-i$, we show that the difference between the two cells is at least $\frac{1}{4i}$ with constant probability. We do this by case analysis. In the first case, exactly one of the two conditions are true, which practically means at time $n-i$, one of the two neighboring cells is filled, and the other is empty. For the gap between the two to be small, a point close to the filled cell's point $x$ must be sampled. The gap is then at least $\ell$ if all future points lie outside $[x-\ell, x+\ell]$. We use a union bound on the probability that there is a point within the interval. Therefore:

$$\Pr_{\{x_n\},j}\left[|a_j - a_{j+1}| \geq \frac{1}{4i} \;\middle|\; T_j > n-i \oplus T_{j+1} > n-i\right] \geq 1 - i \cdot \frac{2}{4i} \geq \frac{1}{2} \quad (13)$$

The other case is that both of the conditions are true, and thus none of the cells are filled. We then consider when the first of them is filled. Note that $n - \min\{T_j, T_{j+1}\} < i$. Therefore,

$$\Pr_{\{x_n\},j}\left[|a_j - a_{j+1}| \geq \frac{1}{4i} \;\middle|\; T_j > n-i \land T_{j+1} > n-i\right] \geq 1 - \left(n - \min\{T_j, T_{j+1}\}\right) \cdot \frac{2}{4i} > \frac{1}{2} \quad (14)$$

Combining (12) , (13) and (14) we get

$$\Pr_{\{x_n\},j}\left[|a_{j+1} - a_j| \geq \frac{1}{4i}\right] \geq \frac{i}{2n} \quad (15)$$

Using (15) , we can find the expected difference of adjacent cells.

$$
\begin{aligned}
\mathbb{E}_{\{x_n\},j}\big[|a_{j+1}-a_j|\big] &\geq \int_0^1 \Pr_{\{x_n\},j}\big[|a_{j+1}-a_j| \geq y\big]\,dy \\
&> \sum_{i=1}^{n} \int_{\frac{1}{4(i+1)}}^{\frac{1}{4i}} \Pr_{\{x_n\},j}\big[|a_{j+1}-a_j| \geq y\big]\,dy \\
&\geq \sum_{i=1}^{n} \int_{\frac{1}{4(i+1)}}^{\frac{1}{4i}} \Pr_{\{x_n\},j}\left[|a_{j+1}-a_j| \geq \frac{1}{4i}\right]dy \\
&\geq \sum_{i=1}^{n} \int_{\frac{1}{4(i+1)}}^{\frac{1}{4i}} \frac{i}{2n}\,dy \qquad\qquad \text{by (15)} \\
&\geq \sum_{i=1}^{n} \frac{i}{2n}\left(\frac{1}{4i} - \frac{1}{4(i+1)}\right) \\
&= \sum_{i=1}^{n} \frac{i}{2n}\left(\frac{i+1}{4i\cdot(i+1)} - \frac{i}{4i\cdot(i+1)}\right) \\
&= \sum_{i=1}^{n} \frac{i}{2n} \cdot \frac{1}{4i\cdot(i+1)} \\
&= \sum_{i=1}^{n} \frac{1}{8n} \cdot \frac{1}{i+1} \\
&= \frac{1}{8n} \cdot \Omega(\log n).
\end{aligned} \tag{16}
$$

We note that due to the linearity of expectation, this can be used to calculate the expectation of the cost. Formally:

$$
\begin{aligned}
\mathbb{E}_{\{x_n\}}\left[\sum_{j=1}^{n-1}|a_{j+1}-a_j|\right] &= \sum_{j=1}^{n-1}\mathbb{E}_{\{x_n\}}\big[|a_{j+1}-a_j|\big] \\
&= (n-1)\cdot \mathbb{E}_{\{x_n\},j}\big[|a_{j+1}-a_j|\big] \\
&= (n-1)\cdot \frac{1}{8n}\cdot \Omega(\log n) \\
&= \Omega(\log n).
\end{aligned} \tag{17}
$$

## 2.4. Stochastic Online TSP

Stochastic online TSP has not been studied very much. In fact, we are only aware of one algorithm for the problem by Kalavas et al. [8]. Note, however, that the technique which is introduced here can be used in most algorithms that use bucketing as a technique. For practical reasons, the number of buckets should be a power of 2. Therefore, Hu's algorithms could use this technique, though there is no guarantee that the resulting algorithm would be any good.

### 2.4.1. Kalavas's Algorithm

Kalavas et al. [8] generalized their algorithm for stochastic online sorting to the high-dimensional case. The algorithm is essentially the same, with only two differences. Instead of using Aamand's algorithm inside the bucket, Bertram's algorithm is used. Bertram's algorithm could also have been used in the one-dimensional case and would give the same asymptotic results. The other difference is how to split the universe into areas for the different buckets. The universe is now a unit $d$-cube, which must be partitioned into $d$-rectangles, which are then associated with each bucket. They do this in a manner similar to a $k$-d tree. Additionally, they define an order of these $d$-rectangles such that adjacent $d$-rectangles in the order are neighbors in space.

By bounding the cost within each bucket and between points in different buckets for each phase, compared with known bounds on the optimal TSP tour for uniformly chosen input points [10, 11], they show that their algorithm achieves a competitive ratio of $O(\log^2 n)$ with high probability for all dimensions. The bounds essentially say that if you pick uniformly random points in the unit $d$-cube, the cost of the optimal tour is $\Theta\left(\sqrt{d} \cdot n^{1-\frac{1}{d}}\right)$ with high probability. Additionally, the cost is $O\left(\sqrt{d} \cdot n^{1-\frac{1}{d}} \cdot V\right)$ in the adversarial case, where $V$ is the volume of a convex polytope.

Considering every phase individually, the cost between different buckets is bounded by the shape of the $d$-rectangles, as they are neighbors. With $2^j$ buckets this distance is at most $O\left(\sqrt{d} \cdot 2^{-\frac{j}{d}}\right)$. This follows from Lemma 5 , which is proved as part of the analysis of our algorithm. As there are $2^j$ of those boundaries, the cost is $O\left(\sqrt{d} \cdot 2^{(1-\frac{1}{d})j}\right) = O\left(\sqrt{d} \cdot n^{1-\frac{1}{d}}\right)$, which is with high probability within a constant factor of the cost of the optimal tour. This mirrors the cost between the buckets from the one-dimensional case. What remains is the cost inside the buckets. The cost of the optimal tour can be upper bounded by [10]. Considering the fact that there are $\Theta\left(\frac{n}{\log^2 n}\right)$ buckets, the volume of the $d$-rectangle is $\Theta\left(\frac{\log^2 n}{n}\right)$. This results in an upper bound on the optimal cost with high probability of $O\left(\sqrt{d} \cdot n^{-\frac{1}{d}} \log^2 n\right)$. As we use Bertram's algorithm, we get a competitive ratio of $O(\log n)$ as the bucket capacity is $\Theta(\log^2)$. And there are $\Theta\left(\frac{n}{\log^2 n}\right)$ buckets the total cost is with high probability $O\left(\sqrt{d} \cdot n^{1-\frac{1}{d}} \log n\right)$. This means a competitive ratio of $O(\log n)$ per phase or $O(\log^2 n)$ over the entire execution of the algorithm.

The buckets contain points that are not quite uniformly random, but close. The distribution depends on when the buckets which this bucket spans from the previous phase are full. When one is full the points are uniformly distributed in one half of the $d$-rectangle, and when both are full, the points are uniformly distributed. This structure inside the buckets, can likely be solved more efficiently than adversarially. This idea would make the algorithm conceptually similar to Hu's final algorithm. As noted in Section 2.3.3 buckets of capacity $\Theta(\log n)$ suffice. This can also be used in the high-dimensional case to achieve a competitive ratio of $O\left(\log^{\frac{3}{2}} n\right)$ with high probability.

## 2.5. Stochastic Online Sorting with Larger Array

When we have more cells than the number of points, one approach that can be used for phase-based algorithms is to reduce the number of phases. The later phases simply may never happen since all points have been placed before they are reached. We can apply this to Kalavas's algorithm. As described, all buckets in a phase are filled at the latest after the next phase. This means that if we can find some $k$ such that after $k$ phases $n$ items have been placed, then only at most 1 more phase happens, and the remaining phases never happen. It turns out that this becomes $O(1 + \log \varepsilon^{-1})$ phases, which we show in the proof of Theorem 2 . Therefore, a competitive ratio of $O(\log n(1 + \log \varepsilon^{-1}))$ with high probability is achieved. Changing the bucket capacity to $\Theta(\log n)$, we get a competitive ratio of $O(\sqrt{\log n}(1 + \log \varepsilon^{-1}))$. If we try to apply it to any of Hu's algorithms, we encounter a challenge, as we have no guarantee of when the phases are done, so we cannot bound the number of phases easily. In any case, Kalavas's algorithm would still be more efficient in the high probability case, and our algorithm is more efficient in the expected case.

#### 2.5.1. Hashing Algorithm

Another idea is to draw inspiration from hash tables. Stochastic online sorting is similar to hash tables in that the points are sampled uniformly at random in an interval, just as a good hash function provides a random value in an interval. Therefore, if there are $m$ cells, the random points can be rounded down to the nearest multiple of $\frac{1}{m}$ and used as a “hash” value. Using linear probing, the points are placed. Unfortunately, the probing distance becomes quite large when there are only a few cells left. Another problem is that wraparound is very expensive. Even though the points in linear probing terms are close, their actual distance is close to 1. When we have extra space, we can solve both these problems. Firstly, the extra space means a significantly lower expected probing distance [12]. Secondly, by reserving a buffer at the end which we probe into instead of wrapping around, we avoid wraparound with high probability. An expected competitive ratio of $O(1 + \varepsilon^{-1})$ is then achieved [1]. In case of no extra space, [4] states that the cost is roughly $O(\sqrt{n})$ without proof and without source. In [12], it is hypothesized that if the array is filled completely the expected probing distance for each point is $\Theta(\sqrt{n})$. A later edit notes that this has been subsequently proven. Therefore, ignoring wraparound, we get an expected cost of $O(\sqrt{n})$. I have not been able to prove that the wraparounds are few enough to ensure that it does not increase the cost. However, using Markov’s inequality, it can be shown that the cost is at most $O(\sqrt{n} \log n)$, by bounding the probability that a point wraps around depending on its “hash”, which forms a harmonic sum. Note also that this does not mean that the cost is necessarily expected $\Omega(\sqrt{n})$.

# 3. Our Algorithm

The basis of our algorithm is the bucketing technique, the merging technique and the adaptive allocation technique. We recall that Hu was able to achieve $\log n \cdot 2^{O(\log^* n)}$ with these techniques along with recursion. If we could bound the recursion depth to a constant, we could therefore reach our goal of $O(\log n)$, matching the lower bound in online sorting. In order to have a constant depth recursion, the buckets must have constant capacity. This in turn means that recursion is completely unnecessary, as points can be placed however in buckets and the cost is only be higher by a constant factor compared to if the points were placed optimally. We therefore do not recurse.

We present our algorithm for stochastic online TSP. Some simplifications can be made if only stochastic online sorting is concerned. We use $k = \log_2 n - O(1)$ phases, $1, 2, ..., k$. In phase $i$ we have $2^{k-i}$ buckets. Each bucket has an associated $d$-rectangle as a span. Points are placed into the bucket that spans the point. In Hu's algorithms and Kalavas's algorithm, the phase transition happens when the first bucket is full. This is not possible for our algorithm. As the bucket have constant capacity, this happens very early when $n$ is large, and many buckets will be little filled. We instead transition to the next phase, when all buckets are on average filled by a factor $\frac{3}{4}$. This means that some buckets overflow and do not have space for all points, but then the points are instead placed into nearby buckets.

When transitioning, the buckets' capacities are chosen as in the adaptive allocation technique. Unlike the adaptive allocation technique, we consider the cells from the previous phase to be part of the new bucket. Thus, the buckets from the previous phase cease to exist. We may end up in a situation where the number of new cells required is negative. In that case, some of the cells from the previous phase must be given away to other buckets that are lacking cells. This procedure is, in fact, the same as the procedure for placing points, such that the cells are reused in buckets that are nearby.

The goal is that our algorithm places half of the remaining points in each phase, except for the last phase, where all remaining points are placed. Our algorithm is illustrated in Figure 2.

The buckets of phase $i$ consist of some free cells leftover from the previous phase, as well as some fresh cells called $C_i$, which is a contiguous subarray of $A$. As we want to place $\frac{n}{2^i}$ cells in the $i$'th phase except for phase $k$ the number of cells in total for a phase should be $\frac{4}{3} \cdot \frac{n}{2^i}$.

Let $s_i$ denote the ideal capacity of $C_i$. In phase 1, there are no free cells remaining from the previous phase, as there is no previous phase. Therefore,

$$s_1 = \frac{4}{3} \cdot \frac{n}{2} = \frac{2}{3} \cdot n. \tag{18}$$

For phase $2 \leq i < k$, since the number of cells in the previous phase is twice as many and a $\frac{1}{4}$'th of those are free, that is, half of the cells that phase $i$ needs, therefore

$$s_i = \frac{2}{3} \cdot \frac{n}{2^i}. \tag{19}$$

For phase $k$, the remaining cells are used, therefore

$$s_k = n - \sum_{i=1}^{k-1} s_i = n - \frac{4}{3} \cdot \frac{n}{2} - \sum_{i=2}^{k-1} s_i = n - \frac{4}{3} \cdot \frac{n}{2} - \frac{2}{3} \cdot \frac{n}{2} + \frac{2}{3} \cdot \frac{n}{2^{k-1}} = \frac{4}{3} \cdot \frac{n}{2^k}. \tag{20}$$

Unfortunately, $s_i$ are not necessarily integers, so rounding is required. We split $A$ such that the first $\lceil s_1 \rceil$ cells form $C_1$, the next $\lceil s_1 + s_2 \rceil - |C_1|$ cells form $C_2$, and so on. Therefore, $\left| |C_i| - s_i \right| < 1$, as the rounding of each rounds less than 1 in absolute value, and the two roundings round in opposite directions. In other words, $C_i$ is close to having the ideal capacity.

In the $i$'th phase, the $2^{k-i}$ buckets have associated $d$-rectangles as spans. We define the $d$-rectangles associated with each bucket using the **universe tree**. The universe tree is a $k$-d tree in structure. This means that the root node is the unit $d$-cube, and each node has two children. The children are $d$-rectangles which form an equal partition of the parent's $d$-rectangle. This is done by splitting halfway along a dimension. Which dimension the $d$-rectangle is split along is chosen cyclically by level. For each level, the nodes of that level therefore form an equal partition of the universe, the unit $d$-cube. Since the dimension along which it is split cycles, the number of times it is split along two different dimensions differs by at most 1, and therefore the longest side length is at most twice as long as the shortest side length. The universe tree is illustrated in Figure 1.

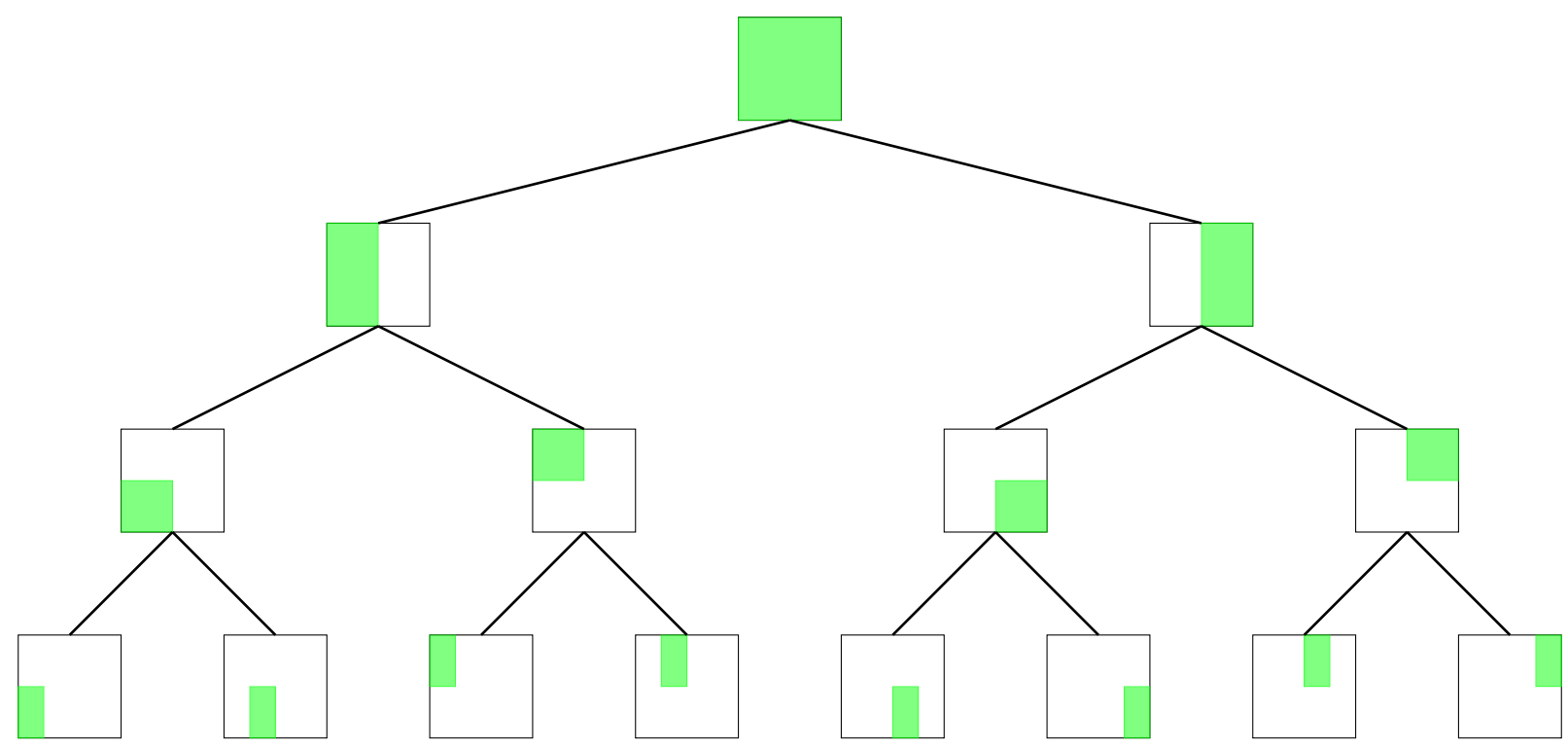

Figure 1: The universe tree in 2 dimensions. The green (2-)rectangles indicate the part of the universe that node is.

In phase $i$, the buckets correspond to the $k - i$'th level of the universe tree, where the root is level 0. We say that a bucket is in the node with the same $d$-rectangle that is the bucket's span. To place a point, the algorithm finds the node in the universe tree, where the $d$-rectangle contains the point, and therefore also the bucket that spans the point. We call this the canonical bucket/node. If there is a free cell in this bucket, the point is placed into the first such cell, which is then filled. If not, we walk to the parent of the canonical node in the universe tree. This node's $d$-rectangle also contains the point. If any of the buckets in this subtree have a free cell, it is placed into one such bucket arbitrarily, otherwise the algorithm conitnues up the universe tree until it succeeds. This never fails as all buckets is in the roots subtree and the algorithm never attempts to place more points than cells in a phase. The exact number of points placed is $\frac{3}{4}$'th of the number of cells rounded down, and in phase $k$, it is the number of cells in the single bucket.

Allocating the buckets in phase 1 is done by splitting $C_1$ into contiguous arrays of equal capacity. Due to rounding, two buckets can differ in capacity by one. Which buckets are large and which are small is chosen arbitrarily. For the remaining phases, the free cells from the previous phase are reused. We start by placing these cells into the new buckets. In the universe tree, this corresponds to moving up every cell one level, since the buckets for this phase are one level above the previous phase. Therefore, for each cell in a bucket, its canonical node is the parent. In case there is not enough space for every cell in a bucket, the insertion procedure is the same as for points described before, i.e., walk up the tree until there is a bucket with remaining capacity. The order in which the cells are placed is arbitrary. Afterwards, cells are taken from $C_i$ to make every bucket equal capacity except for rounding. The cells from $C_i$ are allocated left to right in the universe tree.

Note that a cell may be reused multiple times, and in the extreme case, in every phase.

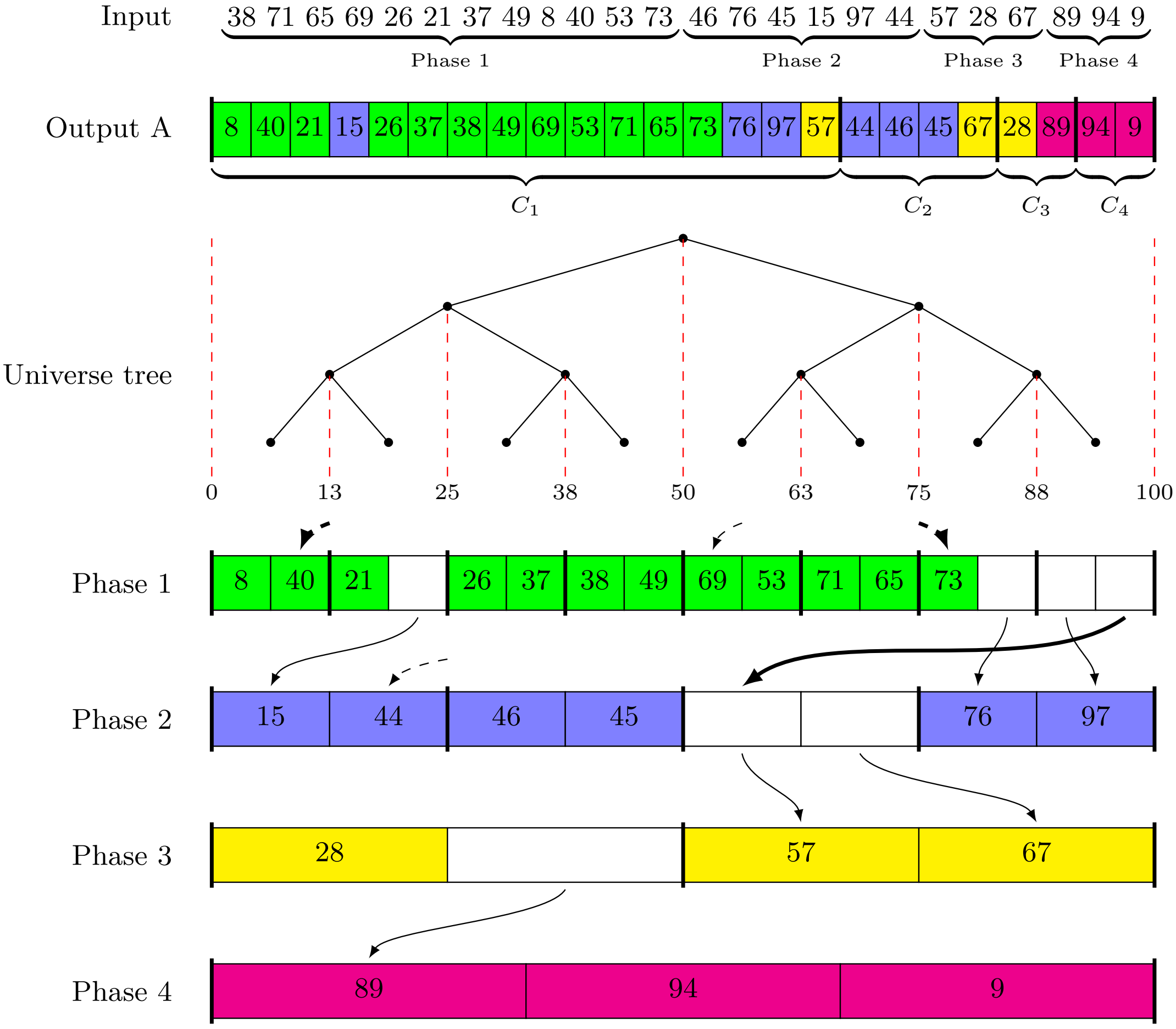

Figure 2: An example of the algorithm with $d = 1$, $n = 24$ points and $k = 4$ phases. We use the distribution of integers from 0 to 99 instead of $[0, 1]$, for example's sake only. The colors indicate the phase a point was placed in. For each node in the universe tree, its span is the interval between the two dashed lines on either side. In the first phase, the buckets correspond to the lowest level in the universe tree, thus spanning intervals $[0, 13), [13, 25), \ldots, [88, 100)$. A dashed arrow above a point indicates that the point was not placed in its canonical bucket. The thickness indicates how far away in tree distance it is placed in terms of the number of levels in the tree traversed. For example, 44 only had to go up one level in the tree, being placed in the $[0, 25)$ bucket instead of its canonical $[25, 50)$ bucket, and thus is thin. However, 40 had to go up two levels in the tree, being placed in the $[0, 13)$ bucket instead of its canonical $[38, 50)$ bucket, and is therefore thick (we allow 40 to be placed in any bucket in the span $[0, 25)$, not necessarily the closest to $[38, 50)$, with a free cell). The arrows between phases indicate the cells that were free at the end of a phase and, therefore, reused in the next phase. The thickness indicates how far from its canonical bucket it was placed. For example, the last cell in phase 1 is not placed in its canonical bucket, which is $[75, 100)$, but in its neighboring bucket, which is $[50, 75)$ and is therefore thick. The remaining cells are taken from $C_i$ from left to right.

# 4. Analysis

Our algorithm is analyzed in three settings. Much of the analysis is the same between the three settings. First, we show some general properties of the algorithm that are used in all three analyses.

In order to analyze the algorithm, we place temporary points in cells until the algorithm places a final point in each cell. In some cases, we replace the temporary points with new temporary points. At the start of the $i$'th phase, we place temporary points in all cells in $C_i$. The idea is that every time a point is replaced due to the triangle inequality, the cost can at most increase by twice the distance between the previous point and the new point. Summing the distances over all replacements and the cost of initial placement in each phase provides an upper bound for the true cost. We upper bound the expected value of this upper bound.

The cost, therefore, comes from the following four sources:

① The cost between different $C_i$.
② The cost for placing temporary points in $C_i$.
③ The cost for placing final points.
④ The cost for replacing temporary points when reusing cells.

For bounding the cost between two points, it is useful to know how far apart the points can be in the worst case, while being inside the same $d$-rectangle in the universe tree.

**Lemma 5.** *The distance between two points in a $d$-rectangle of a node in level $j$ of the universe tree is at most $2\sqrt{d} \cdot 2^{-\frac{j}{d}}$*

*Proof.* The longest distance is between two opposite corners of the $d$-rectangle. Therefore, we bound the side lengths. We know that the volume is $2^{-j}$. The shortest length along a dimension is therefore at most $2^{-\frac{j}{d}}$. Therefore, the longest length along a dimension is at most $2 \cdot 2^{-\frac{j}{d}}$. By the Pythagorean theorem the distance is at most $\sqrt{d \cdot 4 \cdot 2^{-2\frac{j}{d}}} = 2\sqrt{d} \cdot 2^{-\frac{j}{d}}$. □

We consider the cost per phase.

For ① for all but phase 1, when placing temporary points in $C_i$, the first such point is a neighbor of the last in $C_{i-1}$. Since both these points are within the unit cube, we can invoke Lemma 5 with $j = 0$. Therefore, the cost is at most $2\sqrt{d}$.

For ②, we place any point contained in the $d$-rectangle associated with the bucket in every cell of the bucket. For neighboring cells in different buckets, the maximum distance is bounded by any two points of the lowest common ancestor in the universe tree. Note that every node in the universe tree above the level of the buckets is the lowest common ancestor for exactly one neighbor of the buckets. Therefore, we invoke Lemma 5 for every such node. For $d \geq 2$ treating level by level the cost for level $j$ is at most $2^j \cdot 2 \cdot \sqrt{d} \cdot 2^{-\frac{j}{d}} = 2\sqrt{d} \cdot 2^{j(1-\frac{1}{d})} = 2\sqrt{d}2^{(1-\frac{1}{d})j}$. This is a strictly increasing geometric sum with a factor at least $\sqrt{2}$ and therefore asymptotically only the lowest level matters. This is level $2^{k-i}$, and therefore the cost is $O\left(\sqrt{d} \cdot 2^{(k-i)(1-\frac{1}{d})}\right)$. For $d = 1$ the cost does not geometrically increase; however, since the $d$-rectangles are just intervals and they are in increasing order, the distance between two points in adjacent intervals is $2 \cdot 2^{k-i}$. Since there are $2^{k-i}$ of them, the cost is $O(1)$.

For placing a final point, what matters is the difference from the temporary point in the cell. The further up the universe tree the placement goes, the larger that difference may be. Consider the highest node on the path from the canonical bucket to the bucket in which the point was placed in. We call this the placing node. The span of this node spans both the point and the temporary point in the cell. Therefore, this is an upper bound on the difference. Consider marking the placing node for each point. Then the cost is bounded by twice the sum over the products of the number of marks on a node and the maximum distance between two points in its $d$-rectangle. Consider all nodes that are marked, but where no ancestor is marked. Every point placed in this subtree also has a canonical bucket in this subtree, so multiplying the maximum distance between two points in this $d$-rectangle by the number of cells in this subtree is an upper bound on the cost of all those points. For a node to be marked at least one item must have its canonical bucket in one of the child subtrees and the placing bucket in the other. That means that one of the child's subtrees must overflow. Additionally, since none of the ancestors are assumed marked, no items outside this node's subtree are placed in this subtree, therefore, we bound the probability that a child overflows, conditioned on no items being placed with a canonical bucket outside the subtree of the child. By counting all nodes with at least one overflowing child, we upper bound the cost. We do this by upper bounding the probability of overflow and then calculating the expected cost for a node. We use linearity of expectation over all nodes and phases to get an upper bound on the expected cost ③.

As reusing cells uses the same procedure as placing points, the analysis for ④ is the same as for ③, except that calculating the overflow probability is slightly different.

The central argument for the analysis of ③ and ④ is that when we go up one level in the tree, the cost for a node up to quadruples, the higher dimensions the less it increases, as the maximum distance between two points in the $d$-rectangle is at most twice as much and the number of cells in its subtree is twice as many. On the other hand, there are only half as many nodes. We show that the probability of overflow geometrically decreases by a factor of four using a Chernoff bound ( Lemma 9 and Lemma 10 ), thus leading the expected cost per level to decrease by at least a factor of two. As the buckets have a capacity bounded by a constant ( Lemma 6 ), the cost is asymptotically equivalent to the sum of the maximum distances within the $d$-rectangles associated with the buckets, and therefore similar to ② $O\left(\sqrt{d} \cdot 2^{(k-i)(1-\frac{1}{d})}\right)$.

When $d \geq 2$, the cost per phase geometrically decreases, and therefore the cost is asymptotically equivalent to the first phase. Comparing to a bound on the optimal offline cost, with high probability find that the competitive ratio is at most $O(1)$. For $d = 1$, the cost for every phase is asymptotically the same and is $O(1)$. Since there are $O(\log n)$ phases, the cost is $O(\log n)$. With high probability, the optimal cost is $\Omega(1)$, and therefore the expected competitive ratio is $O(\log n)$. When we have a larger array, the number of phases started is simply fewer, therefore the competitive ratio is lower.

We now give a detailed analysis. We show bounds on the number of available cells in each phase, and we show that buckets have a capacity bounded by a constant. Concretely, we show that all buckets have capacity $B$ or $B+1$, where $B = \left\lfloor \frac{n}{3 \cdot 2^{k-2}} \right\rfloor$, and that $B$ is a constant.

**Lemma 6.** $42 \leq B \leq 85$ *and therefore* $B = \Theta(1)$.

*Proof.* From the definition of $B = \left\lfloor \frac{n}{3 \cdot 2^{k-2}} \right\rfloor$ and $k = \lfloor \log_2 n - 5 \rfloor$:

$$
\begin{aligned}
B &= \left\lfloor \frac{n}{3 \cdot 2^{\lfloor \log_2 n - 7 \rfloor}} \right\rfloor \geq \left\lfloor \frac{n}{3 \cdot 2^{\log_2 n - 7}} \right\rfloor \geq 42. \\
B &= \left\lfloor \frac{n}{3 \cdot 2^{\lfloor \log_2 n - 7 \rfloor}} \right\rfloor \leq \left\lfloor \frac{n}{3 \cdot 2^{\log_2 n - 8}} \right\rfloor \leq 85.
\end{aligned} \tag{21}
$$

□

**Lemma 7.** *Let $L_i$ be the cells available in phase $i$. For $1 \leq i < k$, the number of cells available in phase $i$ is bounded by $2^{k-i} \cdot B \leq |L_i| \leq 2^{k-i} \cdot (B+1)$.*

*Proof.* The proof is by induction. For phase 1 we have $L_1 = C_1$, and by definition of $C_1$, $\lfloor \frac{2}{3} \cdot n \rfloor \leq |C_1| \leq \lceil \frac{2}{3} \cdot n \rceil$. Note that $2^{k-1} \cdot B = 2^{k-1} \lfloor \frac{n}{3 \cdot 2^{k-2}} \rfloor \leq \lfloor 2^{k-1} \frac{n}{3 \cdot 2^{k-2}} \rfloor = \lfloor \frac{2}{3} \cdot n \rfloor$ for $k \geq 1$. Similarly $2^{k-1} \cdot (B+1) = 2^{k-1} \left( \lfloor \frac{n}{3 \cdot 2^{k-2}} \rfloor + 1 \right) \geq \lceil 2^{k-1} \frac{n}{3 \cdot 2^{k-2}} \rceil = \lceil \frac{2}{3} \cdot n \rceil$ for $k \geq 1$. Therefore,

$$2^{k-1} \cdot B \leq |L_1| \leq 2^{k-1} \cdot (B+1). \tag{22}$$

For the induction step, recall that in phase $i-1$, $|L_{i-1}|$ cells are available and $\lfloor \frac{3}{4} \; |L_{i-1}| \rfloor$ points are placed. Therefore, there are $\lceil \frac{1}{4} \; |L_{i-1}| \rceil$ free cells which are reused in phase $i$. Therefore, $|L_i| = \lceil \frac{1}{4} \; |L_{i-1}| \rceil + |C_i|$. By definition of $C_i$ for $2 \leq i \leq k-1$, $\lfloor \frac{2}{3} \cdot \frac{n}{2^i} \rfloor \leq |C_i| \leq \lceil \frac{2}{3} \cdot \frac{n}{2^i} \rceil$ we have, similarly to (22) ,

$$2^{k-i-1} \cdot B \leq |C_i| \leq 2^{k-i-1} \cdot (B+1). \tag{23}$$

By the induction hypothesis

$$\begin{aligned} 2^{k-i+1} \cdot B &\leq |L_{i-1}| \leq 2^{k-i+1} \cdot (B+1) \\ 2^{k-i-1} \cdot B &\leq \frac{1}{4} \; |L_{i-1}| \leq 2^{k-i-1} \cdot (B+1) \\ 2^{k-i-1} \cdot B &\leq \left\lceil \frac{1}{4} \; |L_{i-1}| \right\rceil \leq 2^{k-i-1} \cdot (B+1). \end{aligned} \tag{24}$$

As $i \leq k-1$, both bounds are integers, and therefore, after rounding, the inequalities still hold. Combining (23) and (24) completes the proof. □

**Corollary 8**: Every bucket in phase $i < k$ has capacity $B$ or $B+1$.

*Proof.* Following Lemma 7 , the average bucket capacity for a phase is between $B$ and $B+1$. As there is at least one bucket smaller than or equal to the average and at least one larger than or equal to, no bucket can be further from the average than 1. □

We now show that for both placing items and reusing cells, the probability that a node overflows is geometrically decreasing when going up the tree.

**Lemma 9.** *For a node $j \geq 0$ levels above the buckets in the universe tree, the node overflows with probability at most $\left(\frac{1}{4}\right)^j$ when placing points, given that no items are placed with a canonical bucket outside the subtree of the node.*

*Proof.* We use the following Chernoff bound [9, Corollary 4.6] for a binomial distribution $X$ with mean $\mu$:

$$\Pr[X \geq (1+\delta)\mu] \leq e^{-\frac{\delta^2 \mu}{3}}. \tag{25}$$

As each bucket has capacity at least $B$, the node never overflows if at most $2^j B$ items are sampled in this node's span. Therefore, we bound the probability that more were sampled in the phase. Since the expected number of items placed per bucket in a phase is between $\frac{3}{4} \cdot B$ and $\frac{3}{4} \cdot (B+1)$, we know that $2^j \cdot \frac{3}{4} \cdot B \leq \mu \leq 2^j \cdot \frac{3}{4} \cdot (B+1)$. By setting $\delta = \frac{4}{3} \cdot \frac{B}{B+1} - 1 \geq \frac{4}{3} \cdot \frac{42}{43} - 1 = \frac{39}{129}$ using $B \geq 42$ from Lemma 6 we get:

$$
\begin{aligned}
\Pr[X > 2^j \cdot B] &\leq \Pr[X \geq (1+\delta)\mu] \\
&\leq e^{\frac{\delta^2 \mu}{3}} \\
&\leq e^{\frac{39^2 \cdot 2^j \cdot 3 \cdot B}{129^2 \cdot 3 \cdot 4}} \\
&\leq e^{\frac{39^2 \cdot 3 \cdot 42}{129^2 \cdot 3 \cdot 4} 2^j} \\
&\leq e^{\frac{39^2 \cdot 3 \cdot 42}{129^2 \cdot 3 \cdot 4} 2j} \\
&\leq \left(\frac{1}{4}\right)^j.
\end{aligned}
\tag{26}
$$

□

**Lemma 10.** *For a node $j \geq 0$ levels above the buckets in the universe tree, the node overflows with probability at most $\left(\frac{1}{4}\right)^j$ when reusing cells, given that no cells are reused with a canonical bucket outside the subtree of the node.*

*Proof.* We use the following Chernoff bound [9, Theorem 4.5] for a binomial distribution $X$ with mean $\mu$:

$$
\Pr[X \leq (1-\delta)\mu] \leq e^{-\frac{\delta^2 \mu}{2}}. \tag{27}
$$

As each bucket has capacity at least $B$, if there are fewer than $2^j B$ free cells in this subtree, this node never overflows. This means that if more than $2^j\left(\frac{B}{2}+1\right)$ items were sampled in this node's span in the last phase, this node never overflows. Therefore, we bound the probability that fewer items were sampled in this node's span. The expected number of items per bucket in the previous phase is at least $\frac{3}{4} \cdot B$; therefore, $\mu \geq 2^{j+1}\frac{3}{4} \cdot B$. By setting $\delta = \frac{1}{3} - \frac{4}{3B} \geq \frac{19}{63}$, using $B \geq 42$ from Lemma 6 , we get,

$$
\begin{aligned}
\Pr\left[X \leq 2^j\left(\frac{B}{2}+1\right)\right] &\leq \Pr[X \leq (1-\delta)\mu] \\
&\leq e^{-\frac{\delta^2 \mu}{2}} \\
&\leq e^{-\frac{19^2 \cdot 2^{j+1} \cdot 3 \cdot B}{63^2 \cdot 2 \cdot 4}} \\
&\leq e^{-\frac{19^2 \cdot 3 \cdot 42}{63^2 \cdot 2 \cdot 4} 2^{j+1}} \\
&\leq e^{-\frac{19^2 \cdot 3 \cdot 42}{63^2 \cdot 2 \cdot 4} 2j} \\
&\leq \left(\frac{1}{4}\right)^j.
\end{aligned}
\tag{28}
$$

□

We can now bound the expected cost per phase.

**Lemma 11.** *The expected cost for phase $i$ is $O\left(\sqrt{d} \cdot \left(2^{1-\frac{1}{d}}\right)^{k-i}\right)$.*

*Proof.* As discussed earlier the cost ① is $O\left(\sqrt{d}\right)$ and ② is $O\left(\sqrt{d} \cdot \left(2^{1-\frac{1}{d}}\right)^{k-i}\right)$.

For phase $j < k$, consider a node $j$ levels above the buckets. By a union bound, Lemma 9 and Lemma 10 , the probability that a node is placing is at most $2 \cdot \left(\frac{1}{4}\right)^{j-1}$, for $j \geq 1$. The formula also holds for $j = 0$ as it claims a probability more than 1. The cost this node incurs if it is placing is at most the number of cells in its subtree times twice the maximum distance of two points in the $d$-rectangle associated with the node. As there are $2^j$ buckets with at most $B + 1$ cells, and the maximum distance between two points in the $d$-rectangle is $2\sqrt{d} \cdot 2^{-\frac{(k-i)-j}{d}}$ due to Lemma 5 , the expected cost for this node is at most $2 \cdot \left(\frac{1}{4}\right)^{j-1} \cdot 2^j \cdot (B+1) \cdot 2 \cdot 2\sqrt{d} \cdot 2^{-\frac{(k-i)-j}{d}}$. As there are $2^{(k-i)-j}$ nodes $j$ levels above the buckets the expected cost for this level is at most:

$$\begin{aligned} & 2 \cdot \left(\frac{1}{4}\right)^{j-1} \cdot 2^j \cdot (B+1) \cdot 2 \cdot 2\sqrt{d} \cdot 2^{-\frac{(k-i)-j}{d}} \cdot 2^{(k-i)-j} \\ & = 32 \cdot (B+1) \cdot \left(\frac{1}{2}\right)^j \sqrt{d} \cdot \left(2^{1-\frac{1}{d}}\right)^{(k-i)-j} \\ & \leq 32 \cdot (B+1) \cdot \left(\frac{1}{2}\right)^j \sqrt{d} \cdot \left(2^{1-\frac{1}{d}}\right)^{k-i} \\ & = O\left(\left(\frac{1}{2}\right)^j \sqrt{d} \cdot \left(2^{1-\frac{1}{d}}\right)^{k-i}\right). \end{aligned} \tag{29}$$

Summing over all levels form a geometrically decreasing sum and therefore the cost ③ and ④ is

$$O\left(\sqrt{d} \cdot \left(2^{1-\frac{1}{d}}\right)^{k-i}\right). \tag{30}$$

For phase $k$ using Lemma 6 and Lemma 7 , the number of cells in phase $k$ is $|C_k| + \left\lceil \frac{1}{4} |L_{k-1}| \right\rceil \leq |C_k| + \lceil |L_{k-1}| \rceil \leq \left\lceil \frac{2}{3} \cdot \frac{n}{2^{k-1}} \right\rceil + 2(B+1) = O(1)$. As each item can at most differ by $2\sqrt{d}$ compared to the temporary point in both the case of placing items and reusing free cells due to Lemma 5 , the cost ③ and ④ is $O\left(\sqrt{d}\right)$. □

We show the expected cost of the algorithm, while the goal is in fact to show the expected competitive ratio. For the adversarial case, the distinction is significant, as the adversary can choose an input with a very low offline cost to increase the competitive ratio. In the stochastic case, this is not an issue as the probability that the offline cost is high is high. Lemma 12 formalizes this.

**Lemma 12.** *For stochastic online TSP, an algorithm achieving expected cost $c \cdot \sqrt{d} \cdot n^{1-\frac{1}{d}}$, achieves an expected competitive ratio of at most $8c + o(1)$.*

*Proof.* We first show that the optimal offline cost is at least $\frac{1}{8}\sqrt{\frac{d}{2\pi e}} \cdot n^{1-\frac{1}{d}}$ with probability $1 - o\left(\frac{1}{n}\right)$. To show this, we use a concentration for the TSP tour of uniformly chosen points in the unit $d$-cube [11, Chapter 2]. Let $T_n$ be a stochastic variable denoting the optimal cost of a TSP tour with uniformly chosen points in the unit $d$-cube, and let $\mu_n$ be the expected cost of a TSP tour with uniformly chosen points in the unit $d$-cube. For some constant c',

$$\Pr[|T_n - \mu_n| \geq t] \leq \begin{cases} e^{-c't^2/\log n} & \text{if } d = 2 \\ e^{-c't^2/n^{1-\frac{2}{d}}} & \text{if } d \geq 3. \end{cases} \tag{31}$$

To use this inequality, we need a lower bound on $\mu_n$. It is known that $\mu_n = \beta_d \cdot n^{1-\frac{1}{d}}$, for a constant $\beta_d$ that depends on $d$, and thus a different constant for each dimension [10]. Bounds on $\beta_d$ were later shown in [13, Theorem 1], however, this theorem only works for $d \geq 4$ and both upper and lower bounds it. By inspecting the proof, we find the following lower bound for $d \geq 2$:

$$\beta_d \geq (\text{Vol } B)^{-\frac{1}{d}}\left(1 - \frac{1}{d}\right) \geq \frac{1}{2}(\text{Vol } B)^{-\frac{1}{d}}, \tag{32}$$

where Vol $B$ is the volume of a $d$-ball with radius 1. By using a Stirling bound, we get that:

$$\beta_d \geq \frac{1}{2}\sqrt{\frac{d}{2\pi e}}, \tag{33}$$

therefore:

$$\mu_n \geq \frac{1}{2}\sqrt{\frac{d}{2\pi e}}n^{1-\frac{1}{d}}. \tag{34}$$

Let $t = \frac{1}{4}\sqrt{\frac{d}{2\pi e}}n^{1-\frac{1}{d}}$, since $\mu_n \geq 2t$, then if $T_n$ is within $t$ from $\mu_n$, then $T_n$ is larger than $t$, therefore:

$$\Pr[T_n > t] \geq \Pr[|T_n - \mu_n| < t] = 1 - \Pr[|T_n - \mu_n| \geq t]. \tag{35}$$

For $d = 2$ using (32) we get:

$$\begin{aligned} \Pr[T_n > t] &\geq 1 - e^{-c' t^2/\log n} \\ &= 1 - e^{-c' dn/32\pi e \log n} \\ &= 1 - o\left(\frac{1}{n}\right). \end{aligned} \tag{36}$$

For $d \geq 3$ using (32) we get:

$$\begin{aligned} \Pr[T_n > t] &\geq 1 - e^{-c' t^2/n^{1-\frac{2}{d}}} \\ &= 1 - e^{-c' dn/32\pi e} \\ &= 1 - o\left(\frac{1}{n}\right). \end{aligned} \tag{37}$$

However, all of this is a TSP tour, which is longer than the optimal TSP path. However, the optimal TSP path is at least half as long, due to the triangle inequality, thus at least $\frac{t}{2} = \frac{1}{8}\sqrt{\frac{d}{2\pi e}}n^{1-\frac{1}{d}}$ with probability $1 - o\left(\frac{1}{n}\right)$ as desired.

For $d = 1$, note that $\frac{1}{8}\sqrt{\frac{d}{2\pi e}}n^{1-\frac{1}{d}} \leq \frac{1}{2}$. We show that the cost is at least $\frac{1}{2}$ with probability $1 - o(1)$. If at least one point is in the interval $\left[0, \frac{1}{4}\right]$ and at least one point is in the interval $\left[\frac{3}{4}, 1\right]$, then the cost is at least $\frac{1}{2}$. We therefore bound the probability that this is not the case. The probability that none of the points is in the interval $\left[0, \frac{1}{4}\right]$ is $\left(\frac{3}{4}\right)^n$. Same for the other interval. The probability that at least one happens is at most $2 \cdot \left(\frac{3}{4}\right)^n = o\left(\frac{1}{n}\right)$ by a union bound.

Now that we have high probability concentration bounds on the optimal path length, we can show the lemma. Let $X$ be a continuous random variable denoting the cost of the algorithm and $Y$ a continuous random variable denoting the optimal cost in the offline case. Note that $X$ and $Y$ are dependent variables. The competitive ratio can never be more than $n - 1$, as the distance between two adjacent points cannot be more than the optimal offline cost, i.e., $\frac{X}{Y} \leq n - 1 < n$. Let $r = \frac{1}{8}\sqrt{\frac{d}{2\pi e}}n^{1-\frac{1}{d}}$. By the law of the unconscious statistician [14, Equation 5.19], the expected competitive ratio can be calculated as

$$
\begin{aligned}
\int_{-\infty}^{\infty}\int_{-\infty}^{\infty}\frac{x}{y}f_{X,Y}(x,y)dydx &= \int_{-\infty}^{\infty}\int_{r}^{\infty}\frac{x}{y}f_{X,Y}(x,y)dydx + \int_{-\infty}^{\infty}\int_{-\infty}^{r}\frac{x}{y}f_{X,Y}(x,y)dydx \\
&< \frac{1}{r}\int_{-\infty}^{\infty}x\int_{r}^{\infty}f_{X,Y}(x,y)dydx + n\int_{-\infty}^{\infty}\int_{-\infty}^{r}f_{X,Y}(x,y)dydx \\
&< \frac{1}{r}\int_{-\infty}^{\infty}xf_X(x)dx + n\int_{-\infty}^{\infty}\int_{-\infty}^{r}f_{X,Y}(x,y)dydx \\
&\leq 8c + n\cdot o\left(\frac{1}{n}\right) \\
&= 8c + o(1)
\end{aligned} \tag{38}
$$

□

## 4.1. Stochastic Online Sorting

**Theorem 1.** *There exists an algorithm for stochastic online sorting achieving an expected competitive ratio of $O(\log n)$.*

*Proof.* Note that the cost per phase is $O(1)$ by Lemma 11 when $d = 1$. As there are $k = O(\log n)$ phases, the cost is $O(\log n)$. Lemma 12 finishes the proof. □

## 4.2. Stochastic Online Sorting with Larger Array

**Theorem 2.** *There exists an algorithm for stochastic online sorting with $\lceil (1+\varepsilon)n \rceil$ cells achieving an expected competitive ratio of $O(1 + \log \varepsilon^{-1})$.*

*Proof.* Let $m = \lceil (1+\varepsilon)n \rceil$. The expected cost per phase is $O(1)$ by Lemma 11 , when $d = 1$. Therefore, we only need to count the number of phases started before all points are placed to bound the expected cost. If $\varepsilon \geq 2$, then $|C_1| \geq 2n$ and the number of points that can be placed in the first phase is at least $\left\lfloor \frac{3}{4} \cdot 2n \right\rfloor \geq n$. There is therefore only one phase. Otherwise, using Lemma 7 , Lemma 6 and assuming $t < k$, the number of items that can be placed after $t$ phases is

$$
\begin{aligned}
\sum_{i=1}^{t}|C_i| - \left\lceil \frac{1}{4}\,|L_t| \right\rceil &\geq \left\lfloor \frac{m}{3}\left(2 + \sum_{i=2}^{t}\frac{1}{2^{i-1}}\right) \right\rfloor - \lceil 2^{k-t-2}(B+1) \rceil \\
&\geq \frac{m}{3}\left(2 + \sum_{i=2}^{t}\frac{1}{2^{i-1}}\right) - 2^{k-t-2}(B+1) - 2 \\
&\geq \frac{m}{3}\left(3 - \frac{1}{2^{t-1}}\right) - 2^{k-t-2}(B+5) && t < k \Rightarrow 2^{k-t-2} \geq \frac{1}{2} \\
&\geq m - \frac{m}{3\cdot 2^{t-1}} - 2^{k-t-1}B && B \geq 42 \text{ by Lemma 6} \\
&\geq m - \frac{m}{3\cdot 2^{t-1}} - 2^{k-t-1}\frac{m}{3\cdot 2^{k-2}} \\
&= m - \frac{m}{3\cdot 2^{t-1}} - \frac{m}{3\cdot 2^{t-1}} \\
&= m - \frac{m}{3\cdot 2^{t-2}}
\end{aligned} \tag{39}
$$

We solve for when $n = \frac{m}{1+\varepsilon}$ points are placed:

$$\begin{aligned} m - \frac{m}{3 \cdot 2^{t-2}} &= \frac{m}{1+\varepsilon} \\ 1 - \frac{1}{3 \cdot 2^{t-2}} &= \frac{1}{1+\varepsilon} \\ -\frac{1}{3 \cdot 2^{t-2}} &= -\frac{\varepsilon}{1+\varepsilon} \\ \frac{1}{3 \cdot 2^{t-2}} &= \frac{\varepsilon}{1+\varepsilon} \\ -(\log_2 3 + t - 2) &= \log_2 \varepsilon - \log_2(1+\varepsilon) \\ t &= \log_2 \varepsilon^{-1} + \log_2(1+\varepsilon) + 2 - \log_2 3 \end{aligned} \tag{40}$$

Therefore, $t = O(1 + \log \varepsilon^{-1})$, where $\varepsilon < 2$. Lemma 12 completes the proof. □

## 4.3. Stochastic Online TSP

**Theorem 3.** *There exists an algorithm for stochastic online Euclidean TSP with $d \geq 2$ achieving an expected competitive ratio of $O(1)$.*

*Proof.* By Lemma 11 the expected cost for phase $i$ is $O\left(\sqrt{d} \cdot \left(2^{1-\frac{1}{d}}\right)^{k-i}\right)$. Note that $2^{1-\frac{1}{d}} \geq \sqrt{2}$ for $d \geq 2$, and therefore the sum over all phases is geometrically decreasing. Therefore, the cost is

$$O\left(\sqrt{d} \cdot \left(2^{1-\frac{1}{d}}\right)^{k}\right) = O\left(\sqrt{d} \cdot \left(2^{k}\right)^{1-\frac{1}{d}}\right) = O\left(\sqrt{d} \cdot n^{1-\frac{1}{d}}\right) \tag{41}$$

Lemma 12 finishes the proof. □

# 5. Experimental Evaluation

We evaluate some of the algorithms from previous papers, we surveyed in Section 2, along with our algorithm, to give some insights into the practical performance of these algorithms.

For each data point, the algorithm is run 10 times, and the reported value is the arithmetic mean of the competitive ratios. We generally compute one point for each power of two. When necessary, we will compute more. We limit the number of data points due to computation time. For online sorting, we use the exact optimal offline value, while for online TSP, we cannot practically find the exact optimal value as this problem is NP-complete [15]. We therefore instead use an approximation algorithm. Rosenkrantz et al. [7] analyzed several polynomial-time approximation algorithms for TSP. One of these is Nearest Insertion Algorithm, which starts with an arbitrary point and defines the tour to be the Hamiltonian cycle consisting of just that point. Then, while there are more points, find the closest point, where the distance is the shortest distance among any of the points chosen so far. Then the point is inserted in the cycle such that the length is minimized. The running time is $O(n^2)$, as there are $n$ iterations, and in each iteration, finding the next point takes $O(n)$ time and inserting it takes $O(n)$ time. The algorithm has an approximation factor of $2 - \frac{2}{n}$ [7, Theorem 3].

Rosenkrantz et al. note that an alternative proof for the theorem can be constructed by building the minimum spanning tree and taking twice the cost. This is because if we consider the longest edge of the optimal tour, then removing it, the remaining path length is at most $1 - \frac{1}{n}$ times the length of the optimal tour. This, however, is also a spanning tree, and therefore the minimum spanning tree (MST) must have cost at most this. We can build a tour by starting at some point in the MST and going through the tree in depth-first search order. Each edge in this tour corresponds to a path in the tree, and all these paths in the tree use every edge exactly twice. Due to the triangle inequality, all edges of the tour are at most as long as the path they correspond to, the cost is at most $2 - \frac{2}{n}$ compared to the optimal [7, Lemma 6]. In fact, the Nearest Insertion Algorithm and the MST algorithm are related, in that the order the points are considered in the Nearest Insertion algorithm is the same order as they are considered in Prim's algorithm [16].

How do we then compute an MST? Several MST algorithms exist, for example, Borůvka's algorithm [17], Kruskal's algorithm [18] and Prim's algorithm [16]. Interestingly, Kruskal mentioned an outline of Prim's algorithm in [18] before Prim released his paper. Prim's contribution was how to do this practically by hand. The problem is that all of them run in time $\Omega(dn^2)$, as they all consist of writing down every edge, of which we have $\binom{n}{2}$, as every pair of points can be connected. Using a binary heap for the priority queue in Prim's algorithm, the running time becomes $O(dn^2 + n^2 \log n)$. Using a Fibonacci [19] heap, we can improve this to $O(dn^2)$. A Fibonacci heap is, however, needlessly complicated, as we can just do exactly what Prim did in his paper. We keep a list of the shortest distances to all points from any point in the MST so far. We find the minimum with a linear scan. When we add a new node, we update all distances in the list. Since each scan for the minimum takes $O(n)$ time, each update takes $O(dn)$, and there are $n$ iterations, we get a running time of $O(dn^2)$.

We have so far not used the fact that our graph has a special structure, and only considered algorithms for general graphs. We can use the fact that all distances are Euclidean distances. We can use the Delaunay triangulation to compute the MST in $O(n \log n)$ time for $d = 2$ [20]. This is done by producing a Delaunay triangulation which has size $O(n)$ and then running any of Borůvka's, Kruskal's or Prim's algorithms. This idea unfortunately does not generalize to good algorithms in higher dimensions, as the Delaunay triangulation can be $\Omega(n^2)$ in size. Though algorithms that run in time $o(dn^2)$ exist for higher dimensions, they are not that much better. Yao [21] showed an algorithm achieving $O\left(n^{2-2^{-(d-1)}}(\log n)^{1-2^{-(d-1)}}\right)$. The algorithm works in a similar way to the Delaunay triangulation-based algorithm for $d = 2$, in that a supergraph of the MST is constructed first. It uses the generalized geographic neighbor (GGN) graph, with different parameters. The GGN-graph is a graph where each point has an edge to the nearest point in a direction defined by an infinite cone defined by a basis. By building the union of the GGN-graph for different cones that together partition the universe, we get a supergraph of an MST.

Our graphs have more structure, as the points are distributed uniformly in the $d$-cube. We can use this to get a better algorithm in expectation. Using the idea from [21], we can build the GGN-graph in expected time $2^{O(d)}n$, when the points are distributed uniformly [22]. Afterwards, Prim's algorithm can be used, achieving an expected running time of $2^{O(d)}n \log n$. It is also possible to achieve a slightly better algorithm by creating the MST differently after the graph reduction and achieve an expected running time of $2^{O(d)}n\alpha(n)$ [23].

Due to the complexity of the algorithm and being exponential in $d$, it is unlikely that we can achieve any practical improvement for instances that are practically runnable, compared to Prim's algorithm. This also means that we cannot do as large experiments for large dimensions.

Christofides [24] improved this approximation ratio to $\frac{3}{2}$ by adding a step using a general minimum matching algorithm after constructing the minimum spanning tree. Since general minimum matching takes time $O(n^3)$, using this approximation would only allow us to make small experiments.

Karp [25] showed that for any $\varepsilon$ there exists a probabilistic approximation algorithm for Euclidean TSP that achieves an approximation factor of $1 + \varepsilon$ in time $O(C_\varepsilon n + n \log n)$, where the constant $C_n$ depends on $\varepsilon$ only, if all points are sampled uniformly and independently in a rectangle in the plane. The algorithm is quite simple. The longest edge of the rectangle is chosen. The rectangle is then split perpendicular to this edge, such that one point is in both rectangles and each rectangle has an equal number of points. The algorithm then recurses on each half. Afterwards, the two tours can be stitched together at the common point. One of the occurrences of the common point can be removed. If the number of points left is less than some threshold $t$, the optimal solution is instead found with e.g. Bellman-Held-Karp. [26, 27]. The threshold $t$ depends on $\varepsilon$. Unfortunately, the paper only provides an asymptotic dependence, such that the approximation ratio is $1 + O\left(t^{-\frac{1}{2}}\right)$. Working through the proof, one finds that $\varepsilon \leq \frac{21}{\sqrt{t-1}}$. The algorithm can be extended to higher dimensions, but the analysis only works for $d = 2$. Experimental evaluation suggests that the approximation factor is, in fact, much lower than the proof bounds it to, and that it works well in higher dimensions as well. This is elaborated, along with an analysis of $\varepsilon$, in Appendix A. Arora [28] extended the idea to when the points are not chosen uniformly random, where he showed that an approximation factor of $1 + \varepsilon$ can be achieved with high probability in time $O\left(n(\log n)^{O\left(\sqrt{d}\varepsilon^{-1}\right)^{d-1}}\right)$.

We choose to use the MST approximation for our experiments, because the approximation factor is relatively low and an exact upper bound is known. The running time is relatively low. We use the Delaunay triangulation based algorithm for $d = 2$ and Prim's algorithm for $d > 2$. For each of the experiments, we normalize the competitive ratio with the analyzed bounds on the competitive ratio. Therefore, if we see a horizontal line, the algorithm matches the analysis exactly.

## 5.1. Adversarial Algorithms

For the adversarial algorithms, they might not exhibit worst-case behavior on random input. We therefore also evaluate them on adversarial input. We use the adversary from the lower bound from [1], which we proved in Section 2.1.2. Note that the adversary is always one-dimensional. We evaluate Aamand's algorithm and Bertram's algorithm. To calculate the MST in Betram's algorithm, we use Prim's algorithm as discussed above.

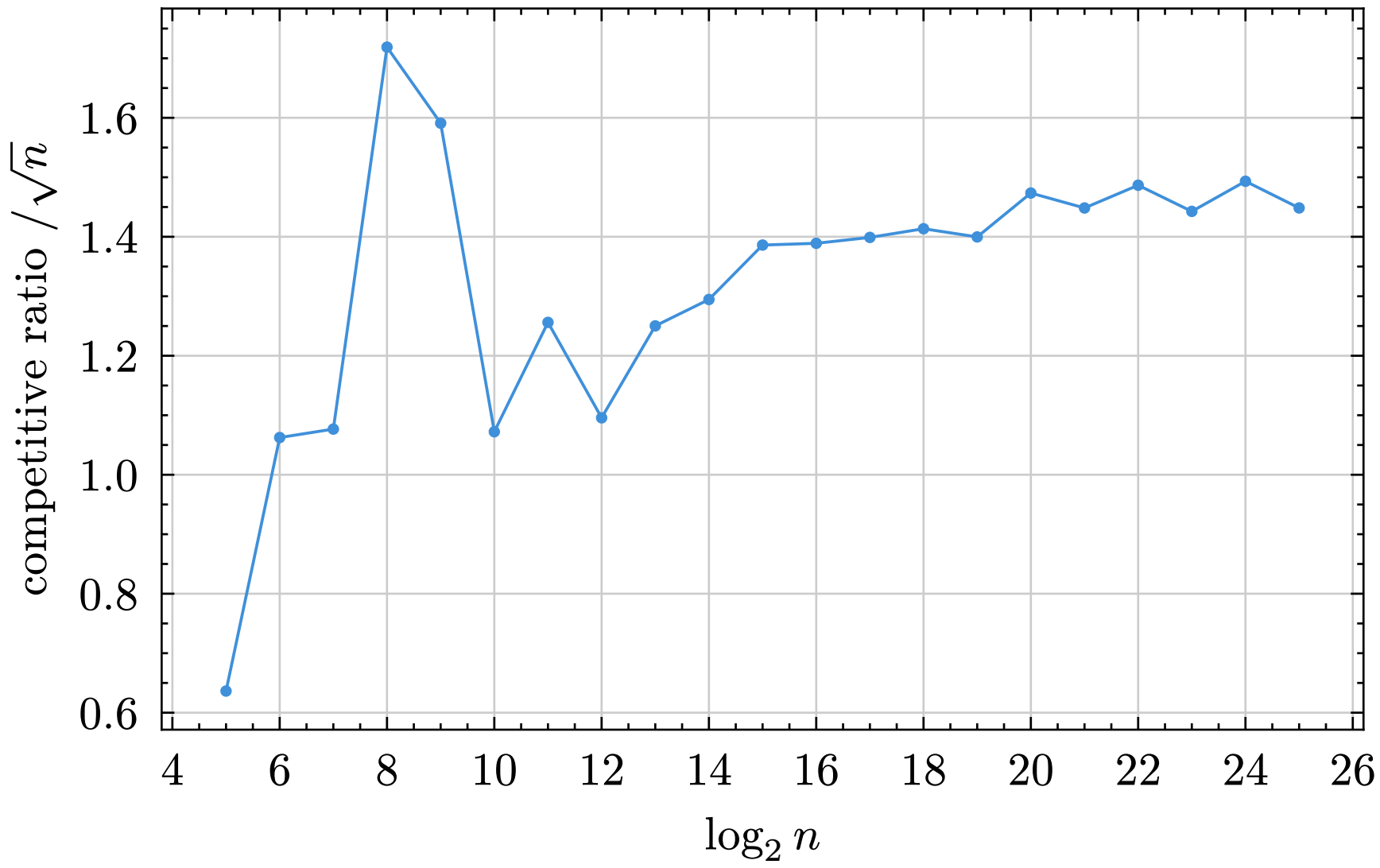

Figure 3: Aamand's algorithm [2] for adversarial input.

As seen in Figure 3, the algorithm does in fact seem to achieve the optimal competitive ratio. We see that the constant seems to go towards somewhere between 1.4 and 1.5.

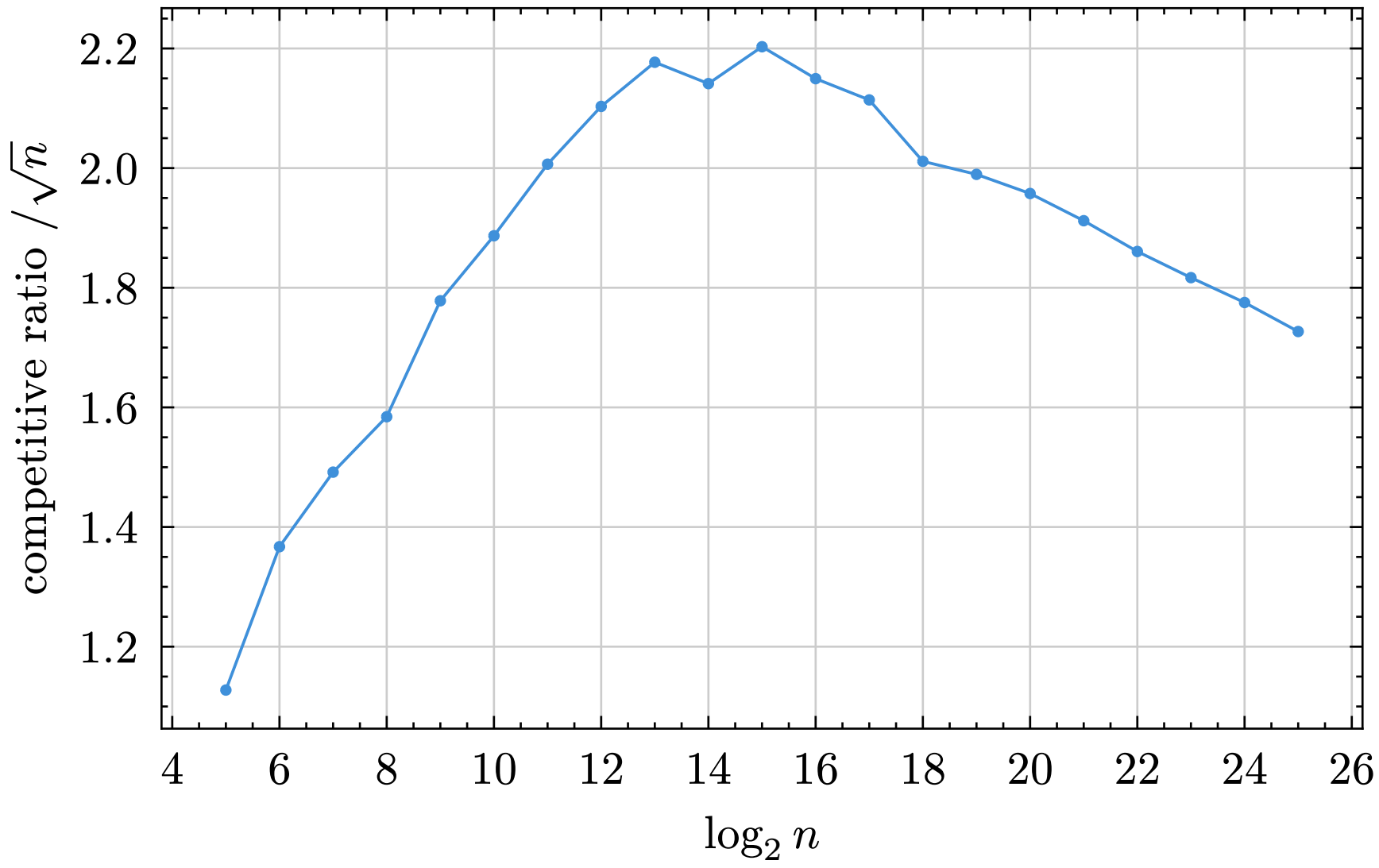

Figure 4: Aamand's algorithm [2] for uniform random input.

The results in Figure 4 indicate that Aamand's algorithm does not exhibit worst-case behaviour on uniform input. The fact that the competitive ratio peaks and then falls is interesting, and we have no explanation for it.

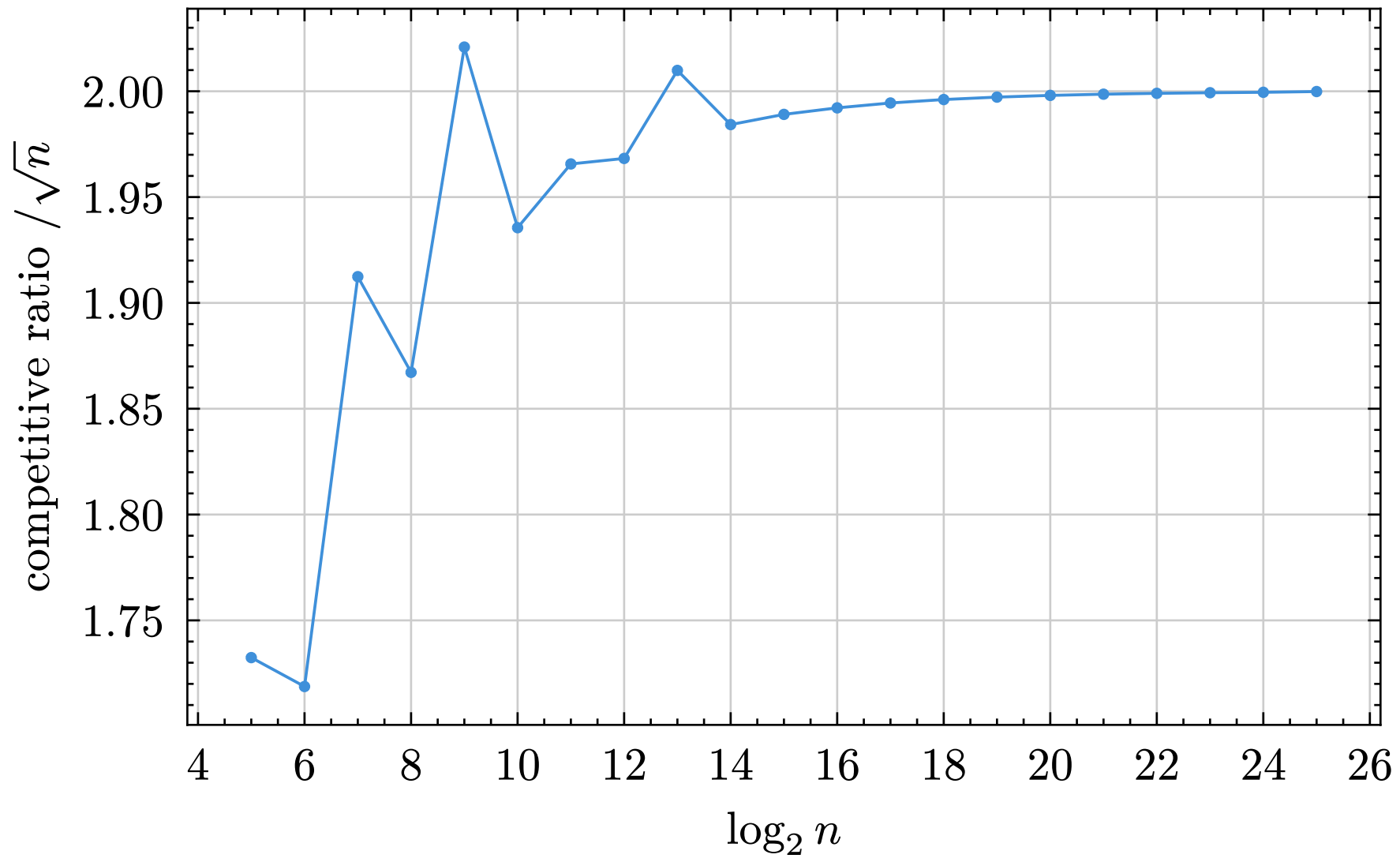

Figure 5: Bertram's algorithm [3] for adversarial input.

In the adversarial case, we can see in Figure 5 that the competitive ratio is clearly $\Theta(\sqrt{n})$, and it is therefore optimal for that adversary. The constant seems to be higher than Aamand's algorithm. This is unsurprising as that algorithm is specifically made for the one-dimensional case.

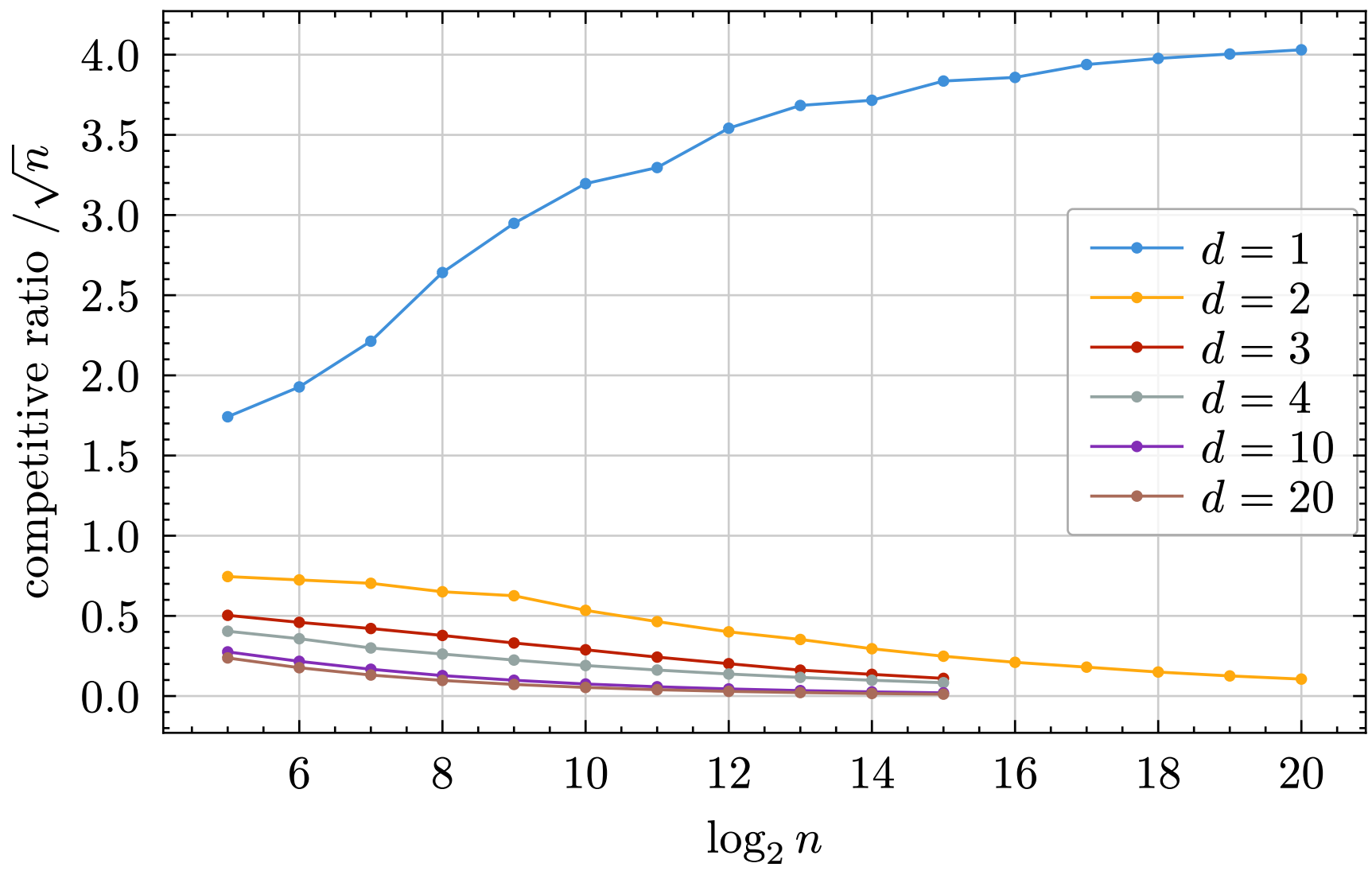

Figure 6: Bertram's algorithm [3] for uniformly random input.

Based on Figure 6, it seems like the competitive ratio is in fact $o(\sqrt{n})$ for $d \geq 2$, when the points are sampled uniformly at random. This is interesting but perhaps not surprising considering the fact that our algorithm also exhibits better performance for $d \geq 2$ than for the one-dimensional case. Unlike Aamand's algorithm, the results for $d = 1$ seem to exhibit worst-case behaviour.

## 5.2. Stochastic Algorithms

Now we evaluate algorithms for the stochastic problem from previous papers. Our algorithm is evaluated later. We evaluate Kalavas's algorithm and Hu's adaptive algorithm. We do not evaluate Hu's merging and Hu's final algorithms, because the parameters required are so large that, with the number of points we can practically evaluate, the merging technique has no effect. For Hu's adaptive algorithm, we do not have any good bounds on the competitive ratio. All we know is that it is subpolynomial. We try to fit a function to the data, thus guessing the complexity. We also evaluate the hashing algorithm.

Kalavas' algorithm uses Aamand's algorithm as a subprocedure in one dimension and Bertram's algorithm otherwise. For simplicity, as the latter also achieves an optimal competitive ratio in one dimension, we always use Bertram's algorithm in the evaluation. This also makes all the dimensions internally comparable, as there is no special case for the one-dimensional case.

In Kalavas' algorithm, we have to set the bucket capacity. The paper only states that the bucket capacity should be $\Theta(\log^2 n)$. We set it to $8 \cdot \lceil \ln^2 n \rceil$, which we found experimentally to be a value high enough such that the probability of failure is low enough, so that it did not happen during our evaluation. The goal is to minimize the bucket capacity, as that improves the cost.

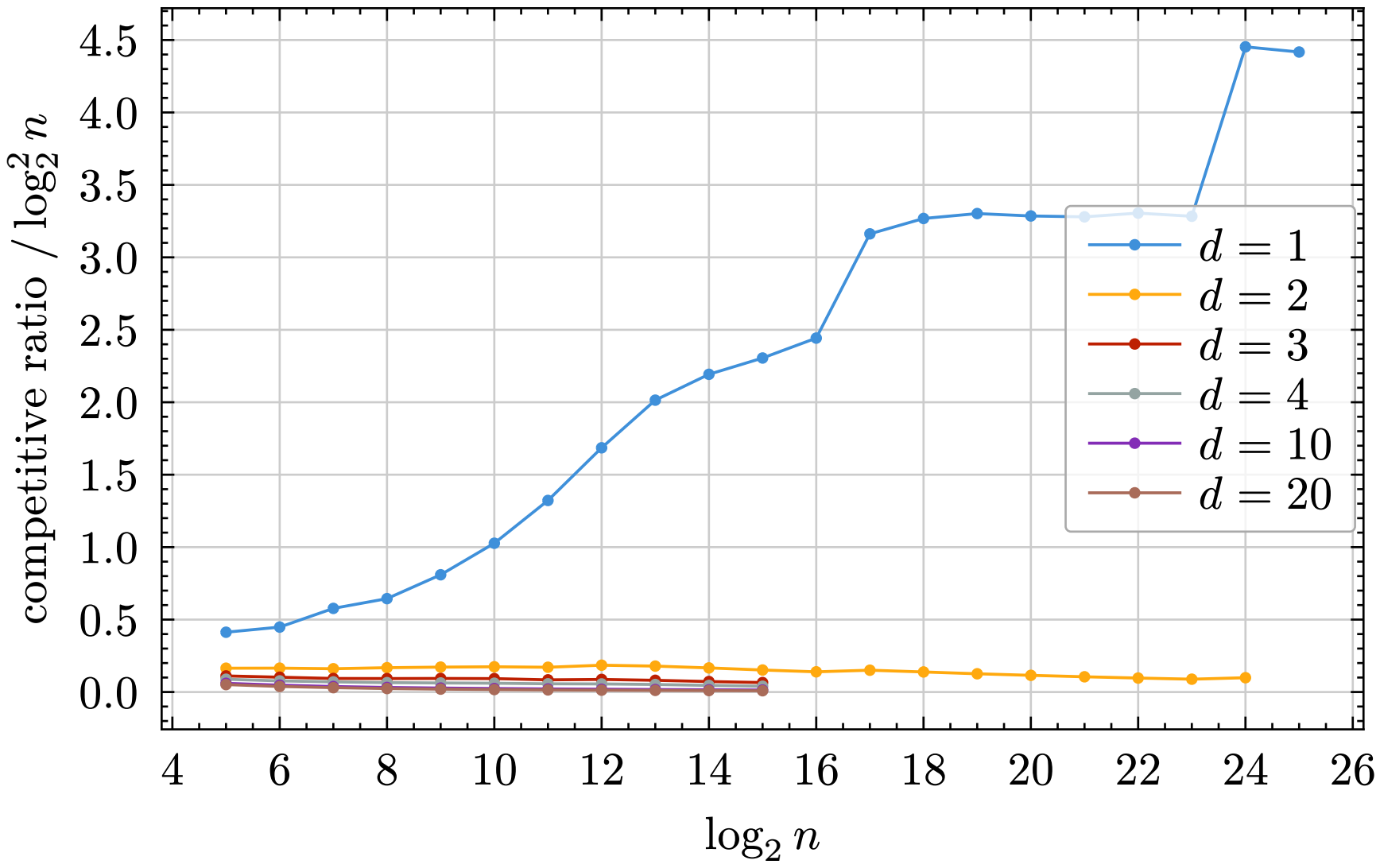

Figure 7: Kalavas's algorithm [8].

Surprisingly, Kalavas' algorithm does not very clearly converge to any point in Figure 7. In particular, the jumps at $n = 2^{17}$ and $n = 2^{24}$ are surprising. To get an idea of what may cause these jumps, we sample a few more data points. We sample points on the form $n = 2^{\frac{k}{10}}$ for $k \in \{50, 51, ..., 225\}$.

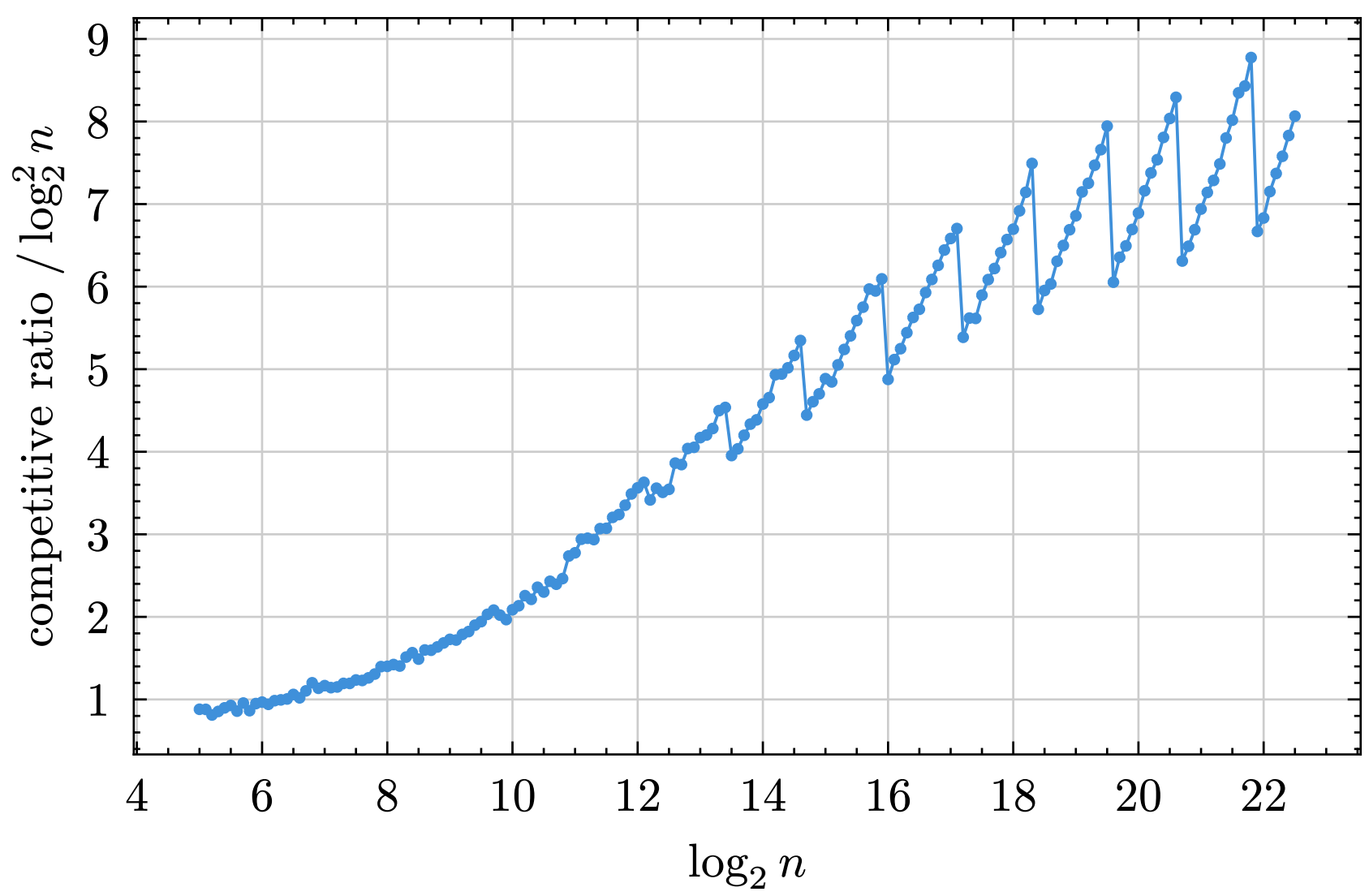

Figure 8: Kalavas's algorithm [8].

As we can see in Figure 8 the competitive ratio looks like a sawtooth, with frequency slightly over one in the log scale. It is therefore reasonable to think that these follow from when the number of phases increases. The two jumps we saw, therefore relates to how the powers of twos interact with more phases.

To see the other dimensions more clearly, we create a plot without $d = 1$.

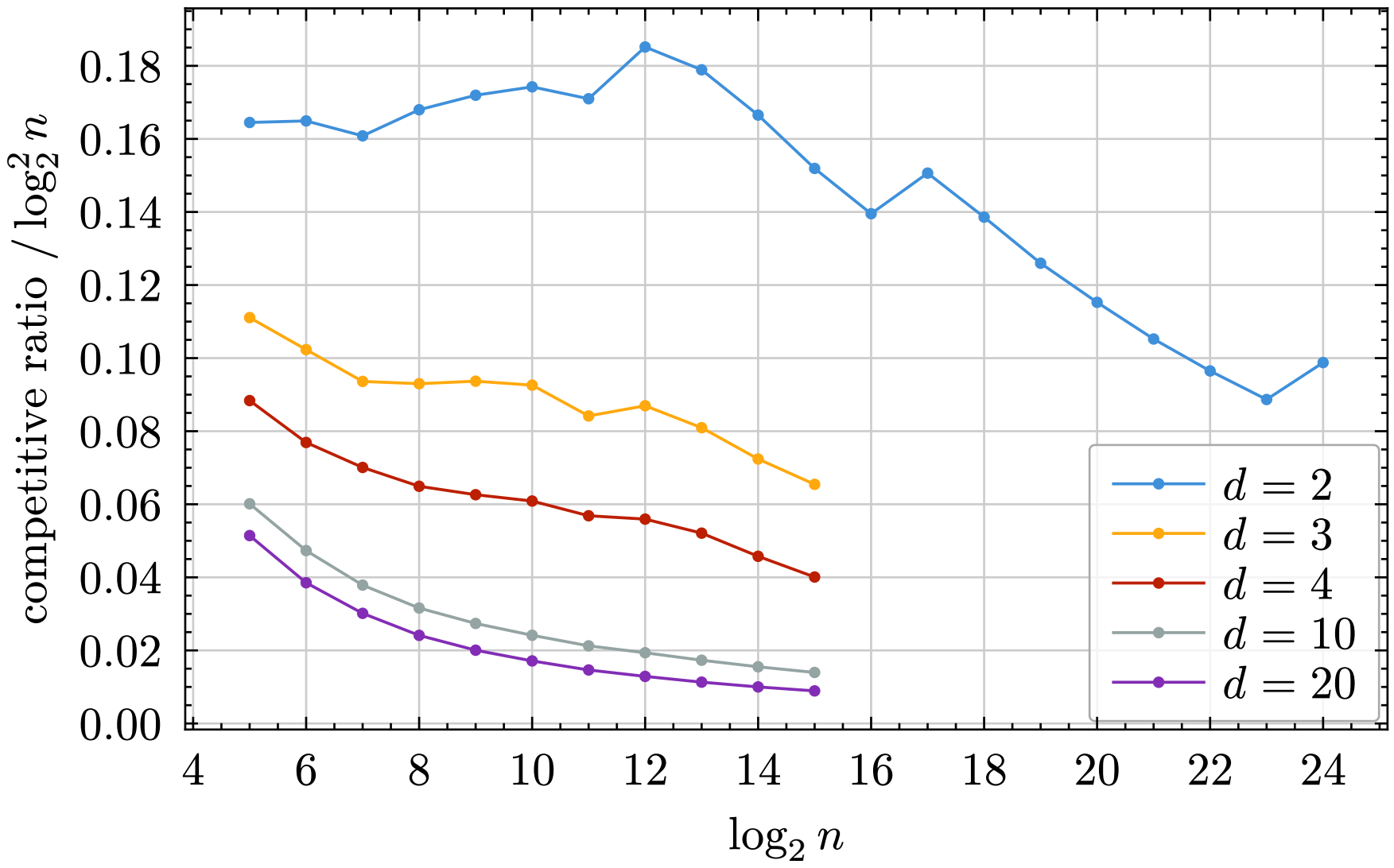

Figure 9: Kalavas's algorithm [8].

We can see in Figure 9 that the jumps in fact also exist at the same spots in $d = 2$, as expected, as the only difference is how to assign points to buckets. We also see that the results seem to be decreasing. This is not surprising as part of the cost comes from the fact that Bertram's algorithm has a worst case competitive ratio of $O(\sqrt{n})$. We saw in Figure 6 that the competitive ratio for Bertram's algorithm seems to be better in expectation when the input is uniformly random. The points are not quite uniformly random in the buckets, but still, the distribution has a lot of structure. A guess is that the competitive ratio is closer to $\Theta(\log n)$ instead, for higher dimensions. We therefore try to plot that.

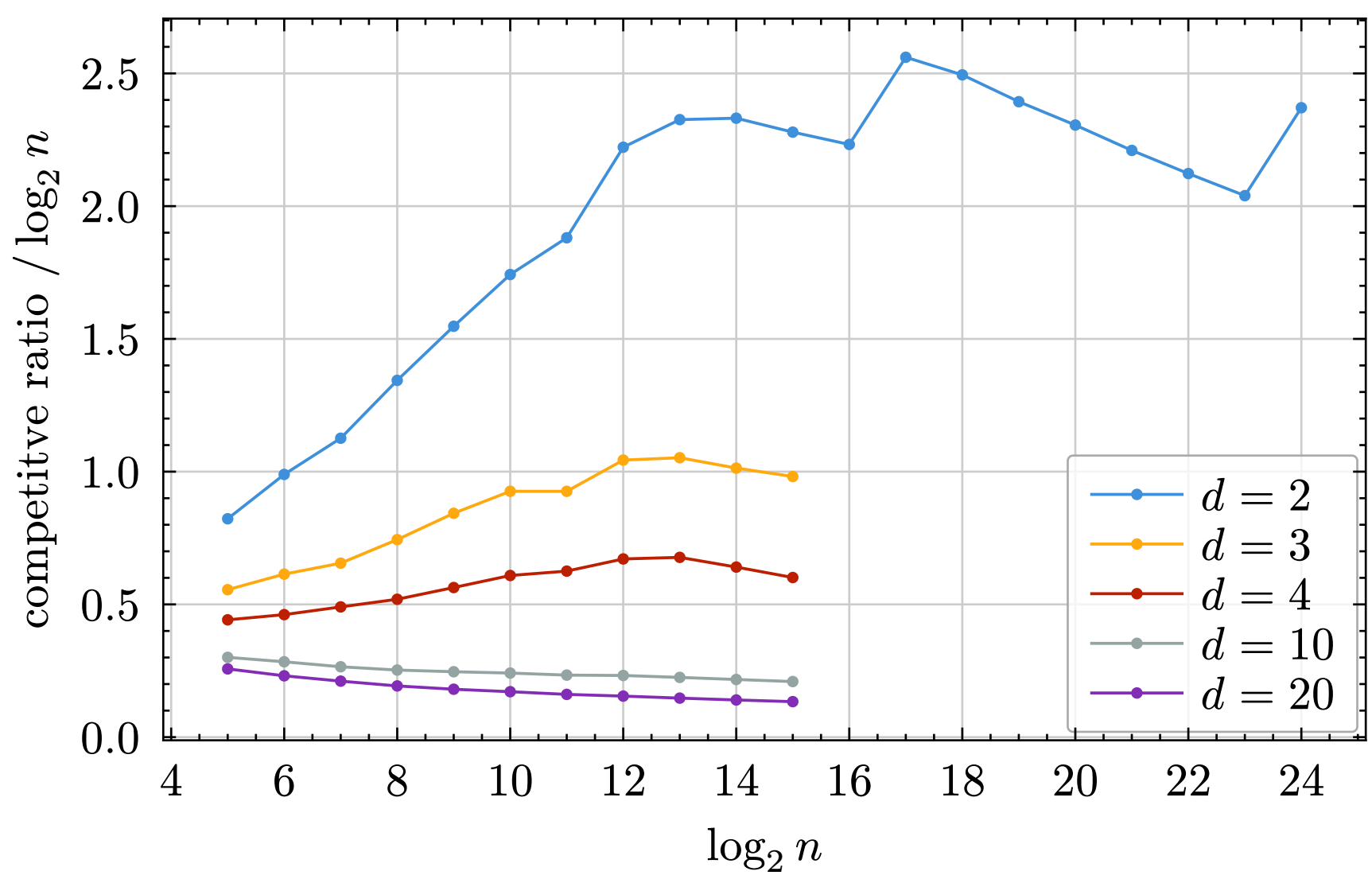

Figure 10: Kalavas's algorithm [8].

The results in Figure 10 are inconclusive. It seems like a better fit, but due to the descending nature, especially for large dimensions, it is plausible that the competitive ratio becomes $o(\log n)$ as $d \to \infty$. It is possible that the cost over phases experiences a decreasing geometric sum with a tighter analysis, although we have not identified such an optimization. That, together with, as stated before, Bertam's algorithm seemingly having a competitive ratio of $o(\sqrt{n})$ in expectation for uniformly random input. It is, however, hard to say when we cannot do evaluations on larger instances.

We showed in Section 2.3.3 that a bucket capacity of $\Theta(\log n)$ was sufficient. We therefore evaluate the algorithm with this change. We noted that it holds for $c \geq 800$. Through experimentation, we found that $c = 200$ was sufficient for it to succeed in the evaluation.

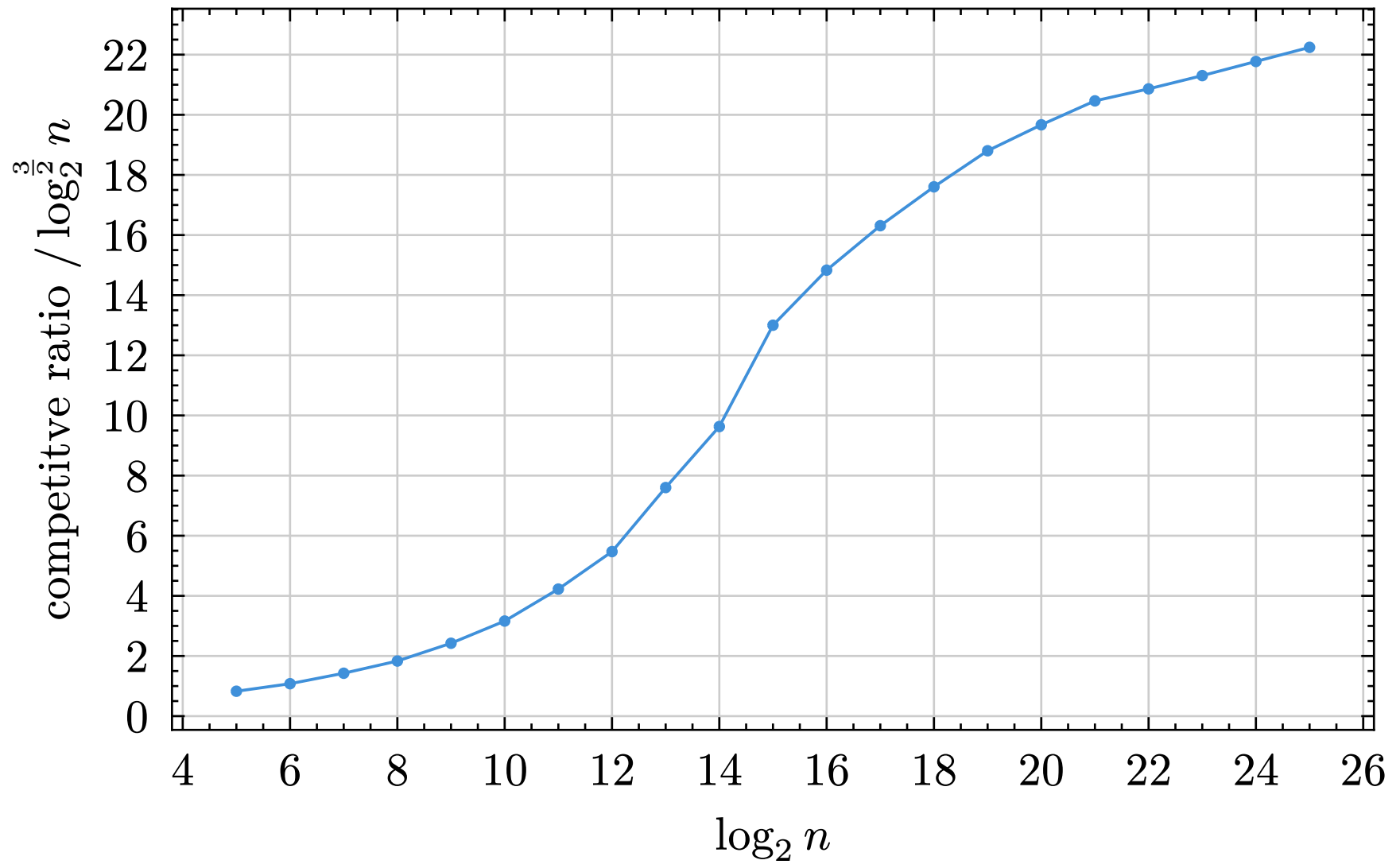

Figure 11: Kalavas's algorithm [8] with $\Theta(\log n)$ bucket capacity for $d = 1$.

It is not clear whether it converges or not. The general shape of the plot looks somewhat like Figure 8 as we would expect.

We now evaluate Hu's adaptive algorithm. As we are guessing on the complexity, we get extra data points to not be as easily deceived by data that happens to fit. By experimenting, we find that $(\log_2 n)^{0.52 \log_2 \log_2 n}$, seems to fit pretty well.

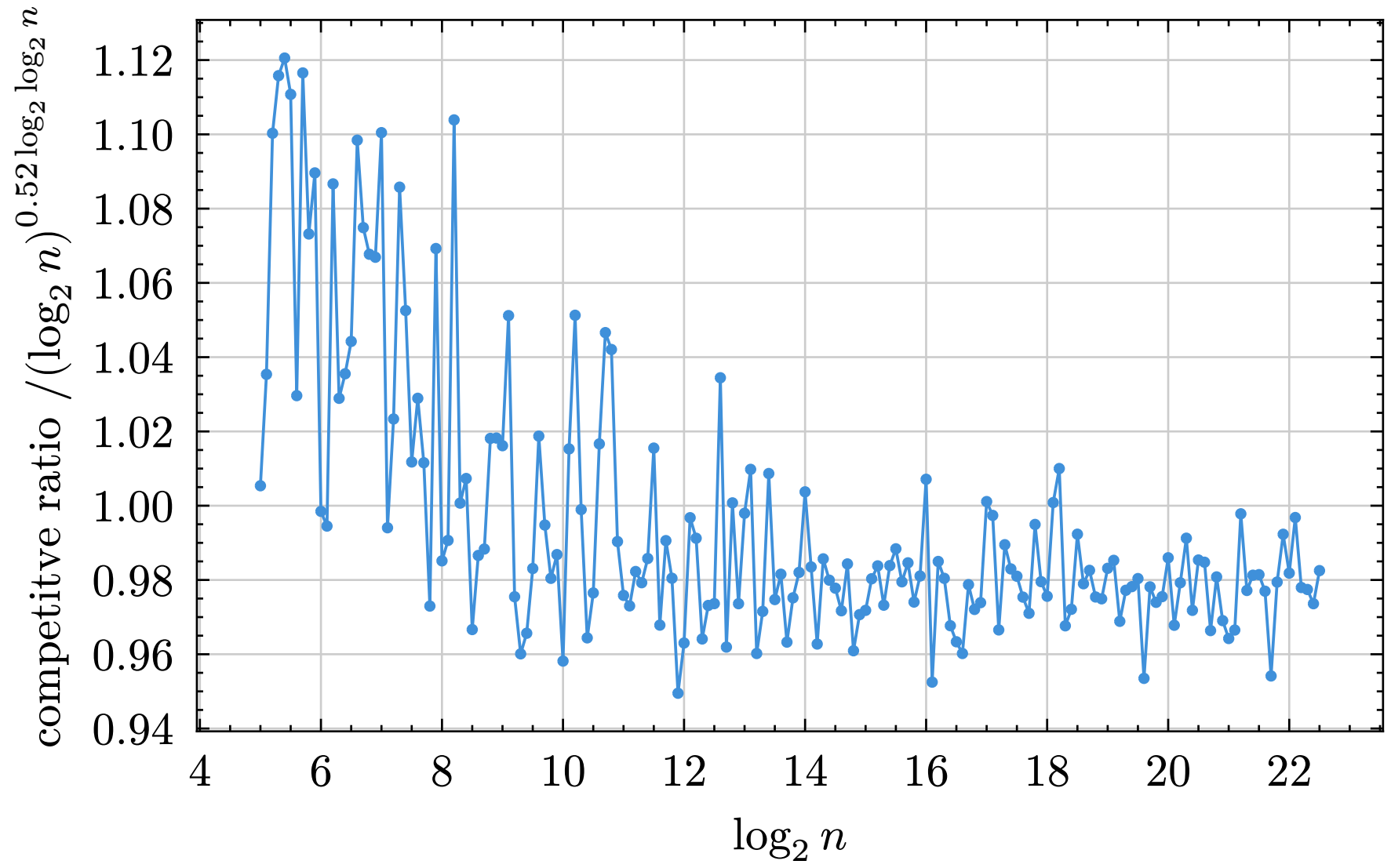

Figure 12: Hu's adaptive algorithm [4].

Of course, we do not know if the cost explodes for larger $n$, but based on the fact that it is horizontal from $2^{12}$ to $2^{22}$, makes it reasonable to think that the guess is not far of.

We now evaluate the Hashing algorithm.

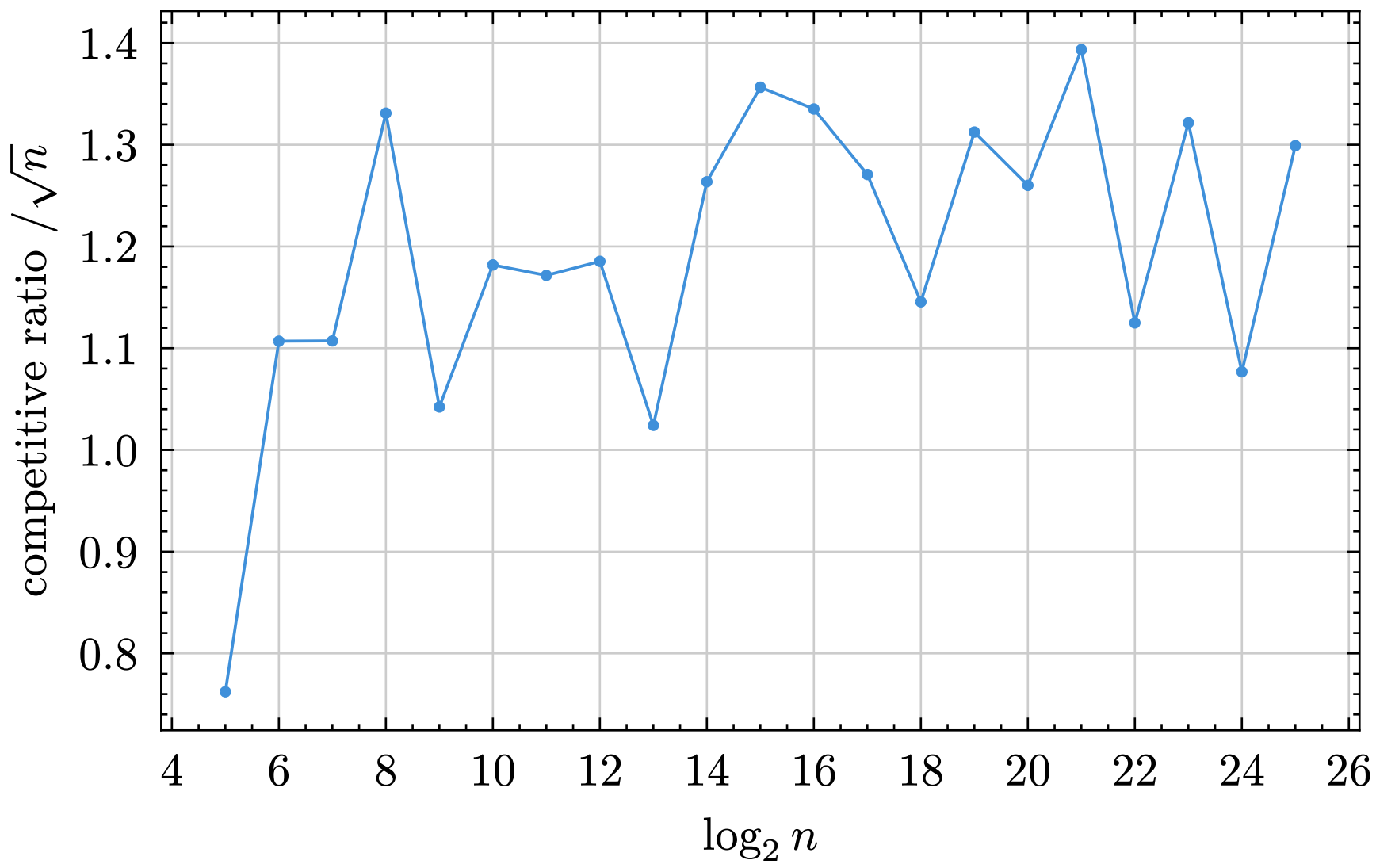

Figure 13: Hashing algorithm [2].

We see in Figure 13 that the algorithm has a very high variance in the competitive ratio, which is not surprising considering the fact that if a point gives considerable cost, it is because it has been probed far. When a point has been probed far, there is a large range of point which would also be probed far. Overall, it does seem like $\Theta(\sqrt{n})$ is the true competitive ratio. Recall that we had no proof of this fact, but had reasons to believe it based on the expected probing distance.

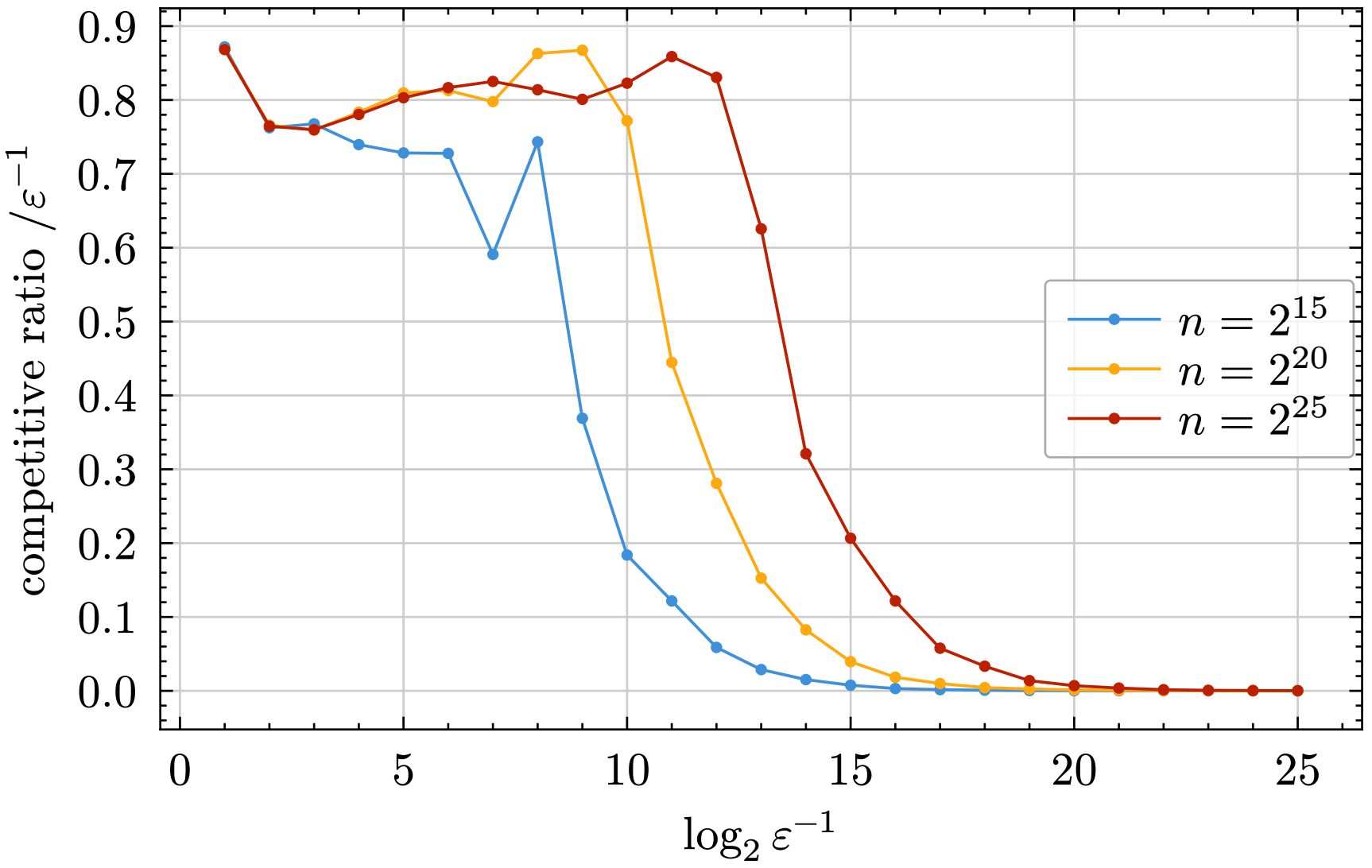

Figure 14: Hashing algorithm [2] with larger array.

Based on Figure 14, we can see that for very large $\varepsilon$ compared to $n$, the cost does actually seem to be $\Theta(\varepsilon^{-1})$; however, when $\varepsilon$ gets smaller, the $\sim \sqrt{n}$ bound for $\varepsilon = 0$ begins to dominate, leading to a much lower cost. We would expect this to happen when $\frac{1}{\varepsilon} = \sqrt{n}$, which aligns with the plot.

## 5.3. Our Algorithm

In our algorithm, the choices of which bucket is inserted into when there are multiple possibilities are arbitrary. In our implementation, we always choose the leftmost bucket. When allocating buckets, we assign capacities such that for each node in the universe tree, the capacity of all buckets in both of the subtrees differ by at most one. We make the left larger than the right. This means that we can set $k$ higher than specified and the analysis still holds if we ignore the lower levels of the universe tree. This practically provides a better result. We evaluate our algorithm in the three settings in which we analyzed it.

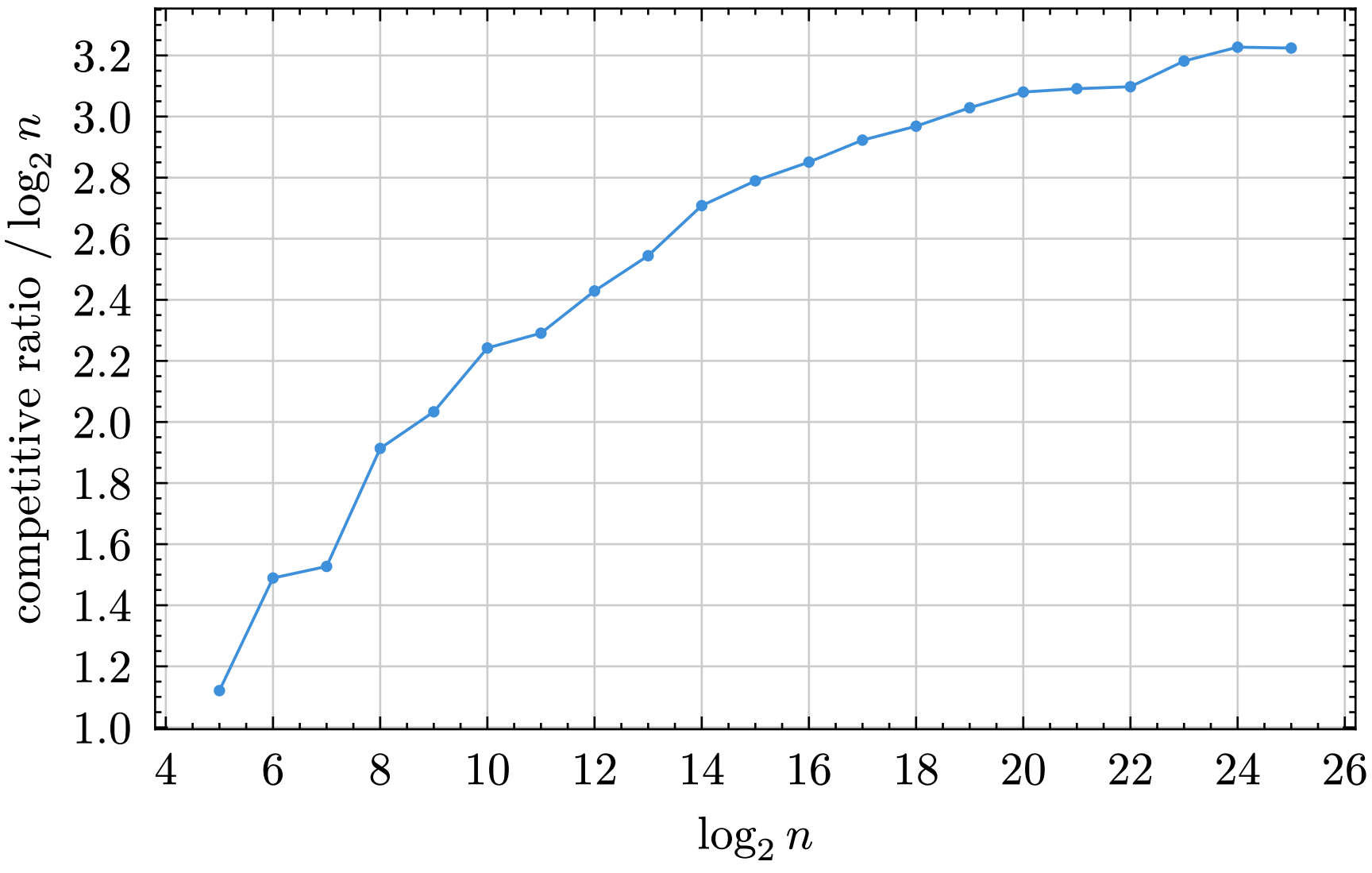

Figure 15: Online sorting

We can see in Figure 15 that it looks like the competitive ratio seems to be stabilizing towards somewhere between 3 and 3.5. The data is inconclusive as we do not have data points for large enough instances. We were not able to evaluate larger instances due to memory requirements.

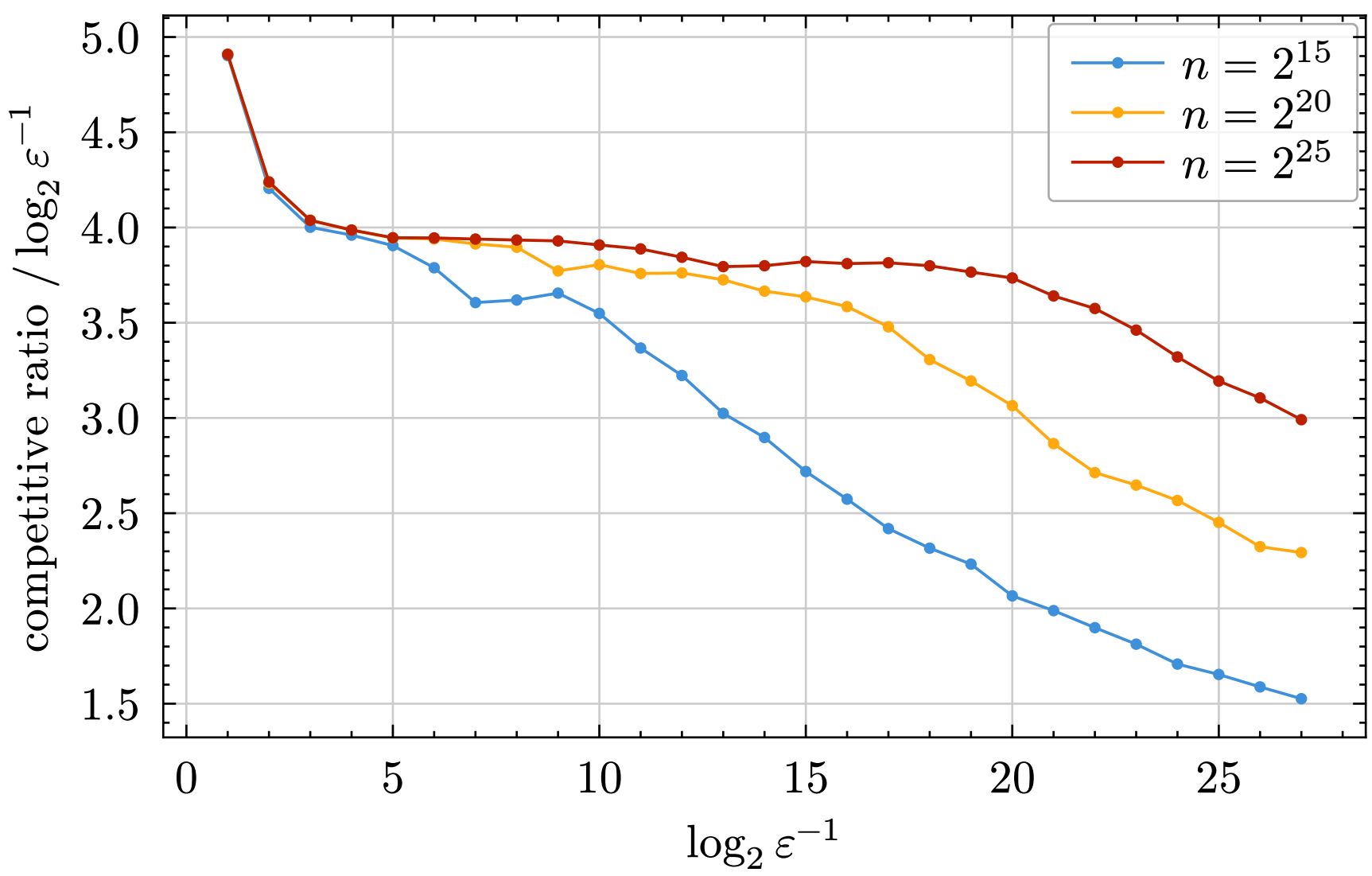

Figure 16: Online sorting with a larger array

For the case with a larger array in Figure 16, we see that the constant seems to be around 4 until $\varepsilon$ becomes small enough compared to $n$ such that $\log n < \log \varepsilon^{-1}$.

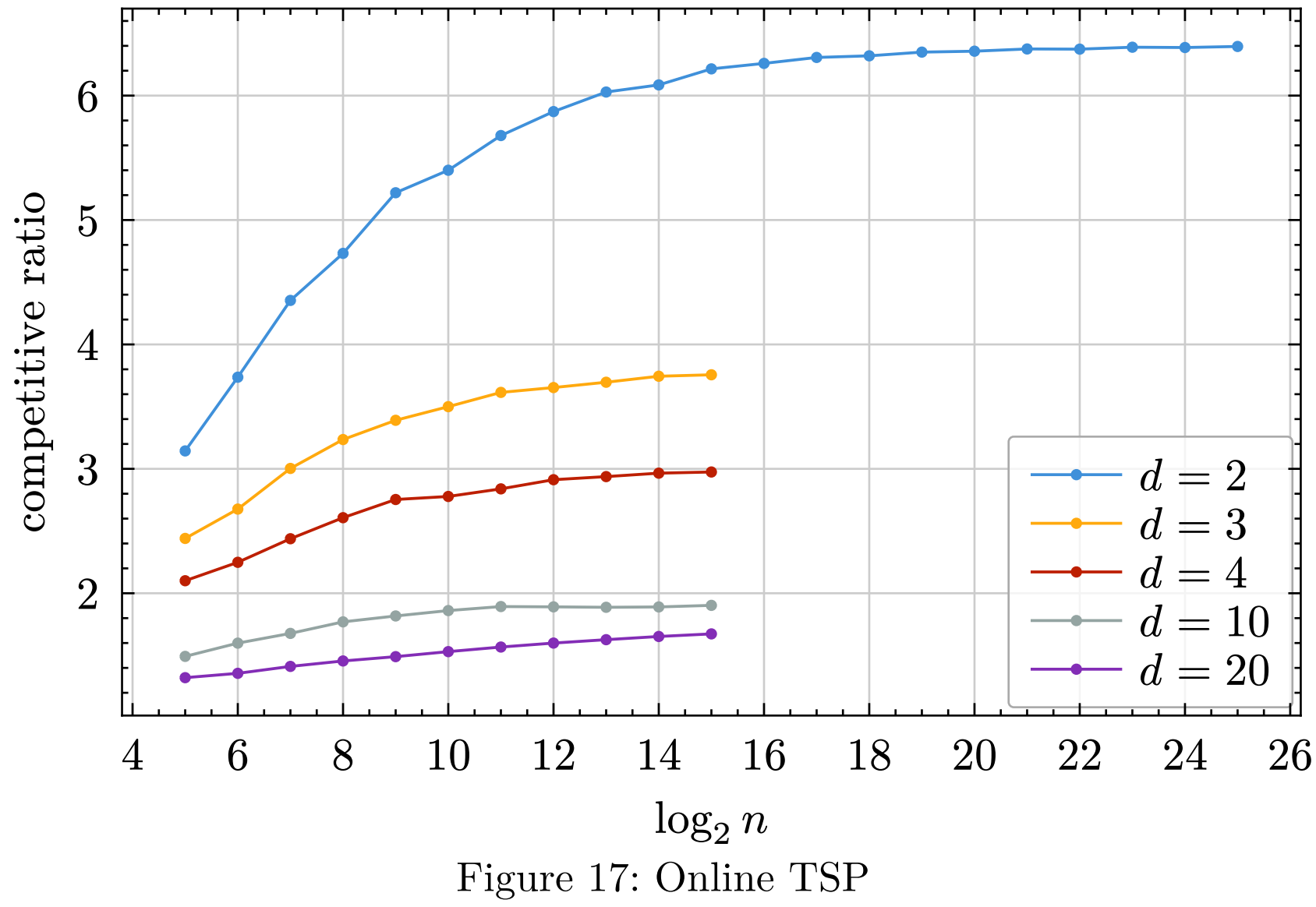

Figure 17: Online TSP

For TSP the competitive ratio is not normalized, as the competitive ratio should be a constant factor. We can see that for different dimensions, it does indeed seem to be constant. Interestingly, this constant is lower for higher dimensions. This is not surprising as the constant which arises from the geometric sum is smaller for larger dimensions (though always between 1 and 3), so this does not explain the full discrepancy between two and twenty dimensions. It should be noted that this may also be a “phantom” effect as the data points are approximated, and the expected quality of this approximation may depend on the dimension.

# 6. Conclusion

We present an algorithm that improves the best known expected competitive ratio for various variants of stochastic online Euclidean TSP. Our algorithm matches the lower bounds in all the presented variants except for one. We also show that our algorithms fare well practically, achieving much better constants than the analysis would suggest. For the setting with the larger array, it would be interesting to find out if there exists a matching lower bound or if a better algorithm exists. We note that we present no improvement in the high probability paradigm. As our algorithm makes heavy use of the linearity of expectation, it seems unlikely that it can be adapted to the high probability case.

In this thesis, we only consider the uniform distribution and the Euclidean distance function. We believe the algorithm can be adapted to other distributions, especially for $d = 1$, such as the normal distribution. The idea here would be to use the inverse cumulative distribution function to build the universe tree such that each bucket has an equal probability of being hit. However, since we have no bounds on the online cost nor the optimal cost, since both can be arbitrarily large, we do not know how to analyze it.

In the adversarial case, our algorithm fails miserably. The worst case competitive ratio is $\Theta(n)$, which can be achieved by picking two opposite corner points of the unit $d$-cube. Then, repeat one of them until half the buckets are full, and then alternate for the rest of the phase. An interesting problem would be to see if an algorithm can be both stochastically and adversarially optimal, or if there potentially is some trade-off between the two.

Our algorithm requires that the points are sampled both uniformly at random and independently. Is it possible to achieve similar bounds if they are not independent? Specifically, our analysis does not hold in that case as the Chernoff bounds require independence. Another question is what if the order is randomized, but the points are chosen ahead of time adversarially, or the points are chosen uniformly at random, but the order can be chosen adversarially. We do not see a way to solve either of these harder problems with our algorithm. Note, though, for the latter, the order within a phase of our algorithm does not matter.

## References

[1] M. Abrahamsen, I. O. Bercea, L. Beretta, J. Klausen, and L. Kozma, "Online Sorting and Online TSP: Randomized, Stochastic, and High-Dimensional," in *32nd Annual European Symposium on Algorithms, ESA 2024, Royal Holloway, London, United Kingdom, September 2-4, 2024*, T. M. Chan, J. Fischer, J. Iacono, and G. Herman, Eds., in LIPIcs. Schloss Dagstuhl - Leibniz-Zentrum für Informatik, 2024, pp. 5:1–5:15. doi: 10.4230/LIPICS.ESA.2024.5 .

[2] A. Aamand, M. Abrahamsen, L. Beretta, and L. Kleist, "Online Sorting and Translational Packing of Convex Polygons," in *Proceedings of the 2023 ACM-SIAM Symposium on Discrete Algorithms, SODA 2023, Florence, Italy, January 22-25, 2023*, N. Bansal and V. Nagarajan, Eds., SIAM, 2023, pp. 1806–1833. doi: 10.1137/1.9781611977554.CH69 .

[3] C. Bertram, "Online Metric TSP," in *33rd Annual European Symposium on Algorithms, ESA 2025, Warsaw, Poland, September 15-17, 2025*, A. Benoit, H. Kaplan, S. Wild, and G. Herman, Eds., in LIPIcs. Schloss Dagstuhl - Leibniz-Zentrum für Informatik, 2025, pp. 80:1–80:9. doi: 10.4230/LIPICS.ESA.2025.80 .

[4] Y. Hu, "Nearly Optimal Bounds for Stochastic Online Sorting," in *Proceedings of the 2026 Annual ACM-SIAM Symposium on Discrete Algorithms, SODA 2026, Vancouver, BC, Canada, January 11-14, 2026*, K. G. Larsen and B. Saha, Eds., SIAM, 2026, pp. 4969–4995. doi: 10.1137/1.9781611978971.180 .

[5] J. Nirjhor and N. Wein, "Improved Online Sorting," in *Approximation and Online Algorithms - 23rd International Workshop, WAOA 2025, Warsaw, Poland, September 18-19, 2025, Proceedings*, J. Matuschke and J. Verschae, Eds., in Lecture Notes in Computer Science. Springer, 2025, pp. 187–197. doi: 10.1007/978-3-032-06706-7_13 .

[6] Y. Azar, D. Panigrahi, and O. Vardi, "Nearly Tight Bounds for the Online Sorting Problem," in *Proceedings of the 2026 Annual ACM-SIAM Symposium on Discrete Algorithms, SODA 2026, Vancouver, BC, Canada, January 11-14, 2026*, K. G. Larsen and B. Saha, Eds., SIAM, 2026, pp. 6642–6658. doi: 10.1137/1.9781611978971.237 .

[7] D. J. Rosenkrantz, R. E. Stearns, and P. M. Lewis, "Approximate algorithms for the traveling salesperson problem," in *15th Annual Symposium on Switching and Automata Theory (swat 1974)*, 1974, pp. 33–42. doi: 10.1109/SWAT.1974.4 .

[8] A. Kalavas, C. Platanos, and T. Tolias, "A Polylogarithmic Competitive Algorithm for Stochastic Online Sorting and TSP," in *43rd International Symposium on Theoretical Aspects of Computer Science, STACS 2026, Grenoble, France, March 9-13, 2026*, M. Mahajan, F. Manea, A. McIver, and K. T. Nguyen, Eds., in LIPIcs. Schloss Dagstuhl - Leibniz-Zentrum für Informatik, 2026, pp. 58:1–58:17. doi: 10.4230/LIPICS.STACS.2026.58 .

[9] M. Mitzenmacher and E. Upfal, *Probability and Computing: Randomized Algorithms and Probabilistic Analysis.* Cambridge University Press, 2005. doi: 10.1017/CBO9780511813603 .

[10] J. Beardwood, J. H. Halton, and J. M. Hammersley, "The shortest path through many points," *Mathematical Proceedings of the Cambridge Philosophical Society*, vol. 55, no. 4, pp. 299–327, 1959, doi: 10.1017/S0305004100034095 .

[11] J. M. Steele, *Probability Theory and Combinatorial Optimization,*. Society for Industrial, Applied Mathematics, 1997. doi: 10.1137/1.9781611970029 .

[12] D. Knuth, “Notes on “Open” Addressing,” 1963. [Online]. Available: https://jeffe.cs.illinois.edu/teaching/datastructures/2011/notes/knuth-OALP.pdf

[13] W. T. Rhee, “On the Travelling Salesperson Problem in Many Dimensions,” *Random Structures & Algorithms*, vol. 3, no. 3, pp. 227–233, 1992, doi: 10.1002/rsa.3240030302 .

[14] H. Pishro-Nik, *Introduction to probability, statistics, and random processes*. Kappa Research LLC, 2014. [Online]. Available: https://www.probabilitycourse.com/

[15] C. H. Papadimitriou, “The Euclidean Traveling Salesman Problem is NP-Complete,” *Theor. Comput. Sci.*, vol. 4, no. 3, pp. 237–244, 1977, doi: 10.1016/0304-3975(77)90012-3 .

[16] R. C. Prim, “Shortest Connection Networks And Some Generalizations,” *Bell System Technical Journal*, vol. 36, no. 6, pp. 1389–1401, 1957, doi: 10.1002/j.1538-7305.1957.tb01515.x .

[17] O. Borůvka, “O jistém problému minimálním,” 1926, Accessed: May 29, 2026. [Online]. Available: https://dml.cz/handle/10338.dmlcz/500114

[18] J. B. Kruskal, “On the Shortest Spanning Subtree of a Graph and the Traveling Salesman Problem,” *Proceedings of the American Mathematical Society*, vol. 7, no. 1, pp. 48–50, 1956, Accessed: May 29, 2026. [Online]. Available: http://www.jstor.org/stable/2033241

[19] M. L. Fredman and R. E. Tarjan, “Fibonacci heaps and their uses in improved network optimization algorithms,” *J. ACM*, vol. 34, no. 3, pp. 596–615, Jul. 1987, doi: 10.1145/28869.28874 .

[20] M. I. Shamos and D. Hoey, “Closest-point Problems,” 1975, doi: 10.1184/R1/6621362.v1 .

[21] A. C.-C. Yao, “On Constructing Minimum Spanning Trees in k-Dimensional Spaces and Related Problems,” *SIAM J. Comput.*, vol. 11, no. 4, pp. 721–736, 1982, doi: 10.1137/0211059 .

[22] J. L. Bentley, B. W. Weide, and A. C.-C. Yao, “Optimal Expected-Time Algorithms for Closest Point Problems,” *ACM Trans. Math. Softw.*, vol. 6, no. 4, pp. 563–580, 1980, doi: 10.1145/355921.355927 .

[23] K. L. Clarkson, “An Algorithm for Geometric Minimum Spanning Trees Requiring Nearly Linear Expected Time,” *Algorithmica*, vol. 4, no. 4, pp. 461–469, 1989, doi: 10.1007/BF01553902 .

[24] N. Christofides, “Worst-Case Analysis of a New Heuristic for the Travelling Salesman Problem,” *Oper. Res. Forum*, vol. 3, no. 1, 2022, doi: 10.1007/S43069-021-00101-Z .

[25] R. M. Karp, “Probabilistic Analysis of Partitioning Algorithms for the Traveling-Salesman Problem in the Plane,” *Math. Oper. Res.*, vol. 2, no. 3, pp. 209–224, 1977, doi: 10.1287/MOOR.2.3.209 .

[26] R. Bellman, “Dynamic Programming Treatment of the Travelling Salesman Problem,” *J. ACM*, vol. 9, no. 1, pp. 61–63, 1962, doi: 10.1145/321105.321111 .

[27] M. Held and R. M. Karp, “A dynamic programming approach to sequencing problems,” in *Proceedings of the 16th ACM national meeting, ACM 1961, USA*, T. C. Rowan, Ed., ACM, 1961, p. 71. doi: 10.1145/800029.808532 .

[28] S. Arora, “Polynomial Time Approximation Schemes for Euclidean Traveling Salesman and other Geometric Problems,” *J. ACM*, vol. 45, no. 5, pp. 753–782, 1998, doi: 10.1145/290179.290180 .

# Appendix

## A. Karp's Partioning Approximation Algorithm

We give an analysis of Karp's parititon approximation algorithm for TSP, where the points are sampled uniformly in a rectangle [25]. The analysis is based on the original paper but avoids asymptotics to give an upper bound on the approximation ratio. The algorithm considers a rectangle of side lengths $a$ and $b$. Assume without loss of generality that $a \leq b$. Let there be $n$ points. If $n \leq t$, an optimal tour is found using an optimal TSP algorithm. We could use Bellman-Held-Karp [26, 27] which runs in time $O(n^2 \cdot 2^n)$. Otherwise, the rectangle is split in two, such that exactly one point is on the border, and there are equally many points in either half. Therefore, the first half has $\lceil \frac{n+1}{2} \rceil$ points and the other half has $\lceil \frac{n}{2} \rceil$ points. We use deterministic select to find the point with rank $\lceil \frac{n+1}{2} \rceil$ with respect to the coordinate of the longer side length of the rectangle. Ties are broken arbitrarily. The algorithm then recurses on each half, returning a tour. These tours are stitched together in the point which overlaps. This point appears twice, so one occurrence can be removed arbitrarily. Due to the triangle inequality, this does not increase the length of the tour. As deterministic select takes time $O(n)$, and we have $O\big(\log\big(\frac{n}{t}\big)\big)$ recursion layers, and therefore $O\big(\frac{n}{t}\big)$ recursion leaves, the running time is deterministically $O\big(nt \cdot 2^t + n\log\big(\frac{n}{t}\big)\big)$. For an expected approximation ratio of $1+\varepsilon$, $t$ can be chosen only depending on $\varepsilon$. Therefore, for a fixed $\varepsilon$, the running time is $O(n \log n)$.

The main insight is that the local optimal tour in a rectangle is only an additive error worse than the part of the optimal global tour that passes through the rectangle. This amount only depends on the length of the perimeter.

Imagine all parts of the optimal tour that pass through the given rectangle. It has $2k$ points where it crosses the rectangle. If we order these points around the perimeter and pairwise connect each neighbour, we get a collection of tours, which all touch the perimeter. We can then connect these tours by going around the perimeter once and connecting each tour once. For the pairwise connections, we have two choices. They use complementary parts of the perimeter, so one of the choices uses at most half the perimeter. This forms a tour which is therefore at most $\frac{3}{2}$ times the length of the perimeter.

We count the sum of the perimeters of all rectangles to bound the error. Imagine that we cyclically decided which axis to cut along instead of the longest. Then, if we have a rectangle of side length $a$ and $b$, then for the first cut we add $2a$ in perimeter as we make one cut of length $a$. In the second cut, we add $2b$ perimeter, as the sum of the lengths of the second axis of the two rectangles is $b$. Then the perimeter length is now $4a+4b$, twice of what we started with. We argue that the strategy of cutting the shortest is no worse than the cyclic strategy. This is because if we make the second cut a long the same axis as the first cut for some of the rectangles, the total length of all cuts is shorter, so the perimeter is increased by less. If the recursion depth is $k$ for some even $k$, the perimeter length is $4 \cdot 2^{\frac{k}{2}}$. If it is odd, then the last makes a locally optimal cut, which is at most half the perimeter we are adding and therefore $4 \cdot 2^{\lfloor \frac{k}{2} \rfloor} \cdot \frac{3}{2} = 4 \cdot 2^{\frac{k}{2}} \cdot \frac{3}{2\sqrt{2}}$. By multiplying by $\frac{3}{2}$ and upper bounding, we get an additive error of at most $9 \cdot \sqrt{2^{k-1}}$.

We now bound the recursion depth $k$. The maximum number of points with a recursion depth of $k$ is $(t-1) \cdot 2^k + 1$. We can therefore calculate $k$ in terms of $n$ and $t$.

$$
\begin{aligned}
n &\le (t-1)\cdot 2^k + 1 \\
\frac{n-1}{t-1} &\le 2^k \\
\log_2 \frac{n-1}{t-1} &\le k.
\end{aligned} \tag{42}
$$

This tells us that if the above is true, then each leaf has at most $t$ points, define:

$$
k = \left\lceil \log_2 \frac{n-1}{t-1} \right\rceil. \tag{43}
$$

We therefore get an error of

$$
9 \cdot \sqrt{2^{\lceil \log_2 \frac{n-1}{t-1} \rceil - 1}} = 9 \cdot \sqrt{2^{\log_2 \frac{n-1}{t-1}}} = 9 \cdot \sqrt{\frac{n-1}{t-1}} \le \frac{9}{\sqrt{t-1}} \cdot \sqrt{n} \tag{44}
$$

Beardwood et al. [10] showed that the optimal path through independently random points in a rectangle is expected $\beta_2 \cdot \sqrt{n}$, as $n \to \infty$. They also stated that $\beta_2 \ge 0.44$. Which means that the expected optimal tour is at most $0.44 \cdot \sqrt{n}$. According to the original paper, the cost tends to $\beta_2 \cdot \sqrt{n}$ for large $n$ with probability 1. Therefore, we can divide our error by the lower bound of the expected cost to get an upper bound for $\varepsilon$:

$$
\varepsilon \le \frac{9}{0.44 \cdot \sqrt{t-1}} \le \frac{21}{\sqrt{t-1}} \tag{45}
$$

Alternatively, we can bound $t$ in terms of $\varepsilon$:

$$
t \le 441\varepsilon^{-2} + 1 \tag{46}
$$

We therefore obtain an expected $(1+\varepsilon)$-approximation, as $n \to \infty$, in time $O\left(n\varepsilon^{-2} \cdot 2^{\varepsilon^{-2}} \cdot n \log(n\varepsilon^{-2})\right)$.

The next question is whether this can be generalized to higher dimensions. The algorithm itself is easily generalized by simply using $d$-rectangles instead of rectangles. The analysis seems harder, though. Bounding the cost of building a local tour based on the optimal global tour is something we do not know how to do. In $d$ dimensions, the running time is $O(dn \log n)$.

We have evaluated Karp's partition approximation algorithm by comparing it with the MST approximation. Like our previous evaluation, we have used an MST-approximation algorithm which underapproximates the tour with a factor at most 2. Therefore, the approximation ratio can be overestimated by a factor of 2. We evaluate it over 10 uniformly generated inputs for each data point, and report the arithmetic mean. As we split by roughly half and we switch to the local optimum when we go below or equal to $t$, only testing powers of two may be misleading, we therefore test with $n = 2^{\frac{k}{10}}$ for $k \in \{30, 31, ..., 150\}$. This should not matter for $t = 2$, as after splitting we always end up with 2 points left exactly, but for higher $t$, we may end up with only $\sim \frac{t}{2}$ points, which affects $\varepsilon$ by $\sim \sqrt{2}$.

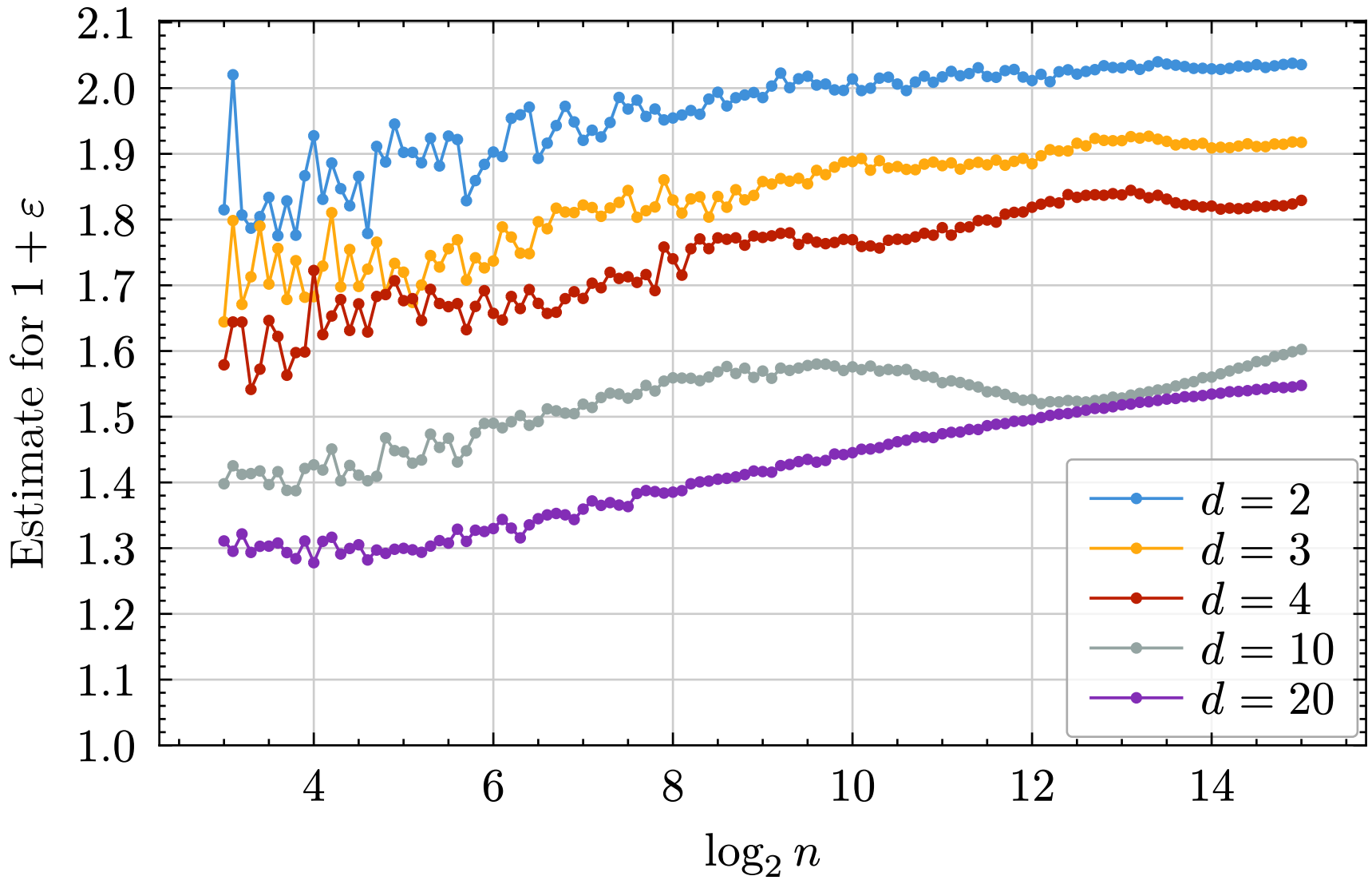

Figure 18: Comparison of Karp's partioning approximation algorithm ($t = 2$) and MST approximation for TSP

We see in Figure 18 that $\varepsilon$ is in fact much lower than the analysis would suggest. This is not surprising as the analysis upper bounds certain steps quite severely. We also see a tendency that higher dimensions actually perform better. This may just be a phantom effect of comparing with MST, as they are all within a factor of 2. We also try with a larger $t$. Setting $t = 10$, according to the analysis, should decrease $\varepsilon$ by a factor of 3 or 67%. We do not expect that much of an improvement as $\varepsilon$ is already way lower than the analysis suggests.

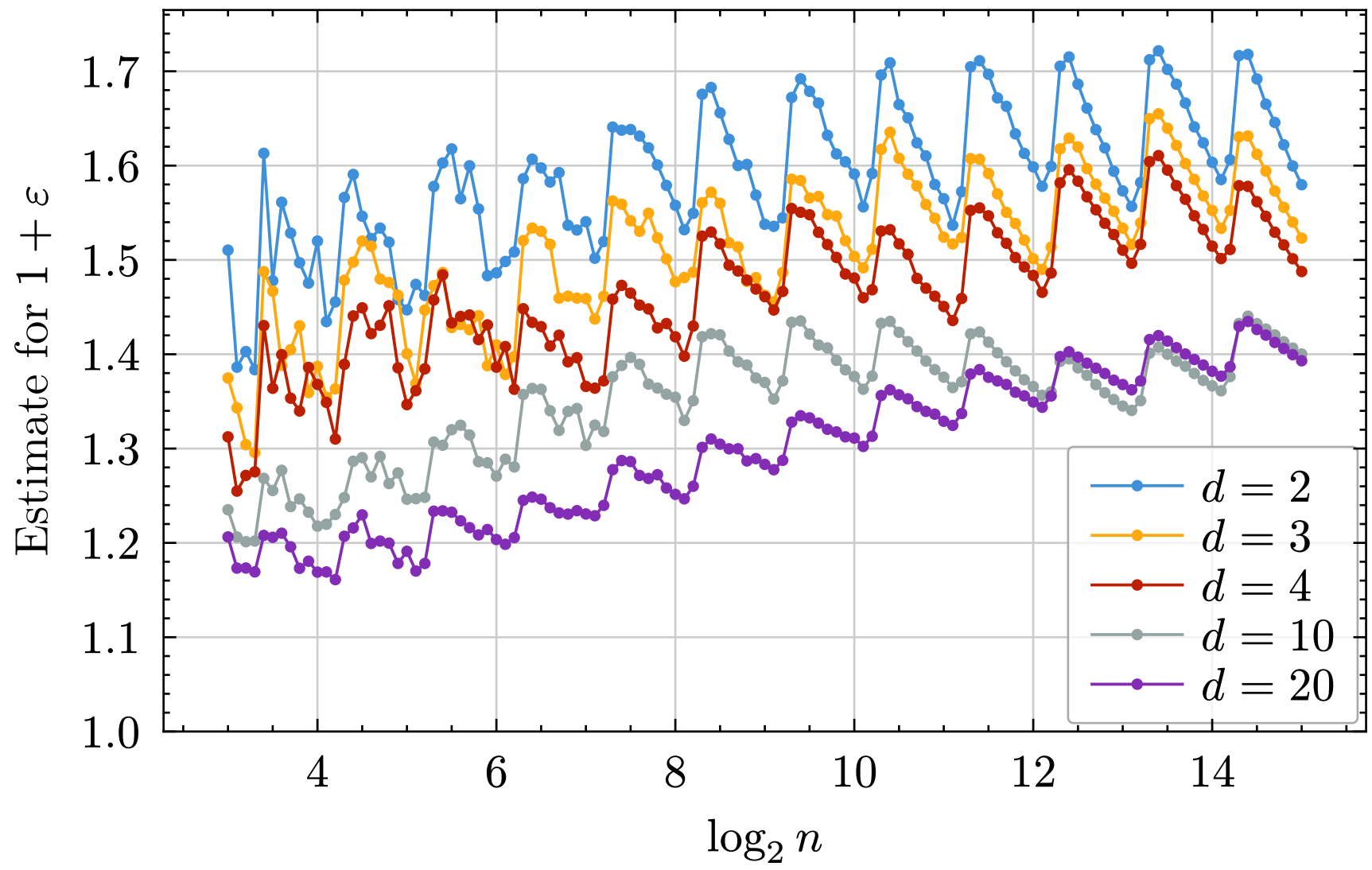

Figure 19: Comparison of Karp's partioning approximation algorithm ($t = 10$) and MST approximation for TSP

As expected, how $n$ is related to an exact power of 2 matters a lot. $\varepsilon$ does not decrease quite by 67%, as seen in Figure 19. Remember that the approximation ratio is $1 + \varepsilon$, so for $d = 2$, $\varepsilon$ was around 1 for $t = 2$ and around 0.6 for $t = 10$, in the points where a threshold of 10 is actually realised, and therefore decreased by $\sim 40\%$. The fact that $t$ is not realisable for all $n$ should also be considered, so it is more fair to say that the decrease is only $\sim 30\%$.